\documentclass[12pt,preprint]{aastex}
\usepackage{amsmath}
\usepackage{hyperref}
\slugcomment{Accepted to {\it The Astrophysical Journal}}
\shortauthors{Mace et al. 2013}
\shorttitle{The T8 Subdwarf WISE~J200520.38+542433.9}

\begin{document}

\def\W2005{WISE~2005+5424}

\title{The Exemplar T8 Subdwarf Companion of Wolf 1130}

\author{Gregory N.\ Mace\altaffilmark{a,b},
J.\ Davy Kirkpatrick\altaffilmark{b},
Michael C.\ Cushing\altaffilmark{c},
Christopher R.\ Gelino\altaffilmark{b},
Ian S.\ McLean\altaffilmark{a},
Sarah E.\ Logsdon\altaffilmark{a},
Edward L.\ Wright\altaffilmark{a},
Michael F.\ Skrutskie\altaffilmark{d},
Charles A.\ Beichman\altaffilmark{b},
Peter R.\ Eisenhardt\altaffilmark{e}, And
Kristin R.\ Kulas\altaffilmark{a}}

\altaffiltext{a}{Department of Physics and Astronomy, UCLA, 430 Portola Plaza, Box 951547, Los Angeles, CA 90095-1547, USA;  gmace@astro.ucla.edu}
\altaffiltext{b}{Infrared Processing and Analysis Center, MS 100-22, California Institute of Technology, Pasadena, CA 91125, USA}
\altaffiltext{c}{Department of Physics and Astronomy, MS 111, University of Toledo, 2801 W. Bancroft St., Toledo, OH 43606-3328, USA}
\altaffiltext{d}{Department of Astronomy, University of Virginia, Charlottesville, VA 22904, USA}
\altaffiltext{e}{NASA Jet Propulsion Laboratory, 4800 Oak Grove Drive, Pasadena, CA 91109, USA}

\begin{abstract}
We have discovered a wide separation (188$\farcs$5) T8 subdwarf companion to the sdM1.5+WD binary Wolf 1130.
Companionship of WISE~J200520.38+542433.9 is verified through common proper motion over a $\sim$3 year baseline.
Wolf 1130 is located 15.83 $\pm$ 0.96 parsecs from the Sun, placing the brown dwarf at a projected separation of $\sim$3000~AU.
Near-infrared colors and medium resolution (R$\approx$2000$-$4000) spectroscopy establish the uniqueness of this system as a high-gravity, low-metallicity benchmark.
Although there are a number of low-metallicity T dwarfs in the literature, WISE~J200520.38+542433.9 has the most extreme inferred metallicity to date 
with [Fe/H] = $-$0.64 $\pm$ 0.17 based on Wolf 1130.
Model comparisons to this exemplar late-type subdwarf support it having an old age, a low metallicity, and a small radius.
However, the spectroscopic peculiarities of WISE~J200520.38+542433.9 underscore the importance of developing the 
low-metallicity parameter space of the most current atmospheric models.
\end{abstract}


\section{Introduction} 
The sample of known T dwarfs has grown considerably since the discovery of Gl 229B \citep{nakajima1995}.
The bulk of the sample has been revealed through surveys like the Two Micron All Sky Survey 
\citep[2MASS;][]{skrutskie2006}, the Sloan Digital Sky Survey \citep[SDSS;][]{york2000}, 
the United Kingdom Infrared Telescope Infrared Deep Sky Survey \citep[UKIDSS;][]{lawrence2007},
the Canada France Hawaii Telescope Legacy Survey \citep[CFHTLS;][]{delorme2008},
and the Wide-field Infrared Survey Explorer \citep[WISE;][]{wright2010}. 
Constraining physical properties like mass, age, and metallicity for isolated field brown dwarfs discovered by such surveys inevitably requires comparison to models.
These models are improved by employing wide-separation T dwarf companions to well studied stars as benchmarks for our analysis of the larger solivagant population
\citep{burgasser2000,burgasser2005,wilson2001,scholz2003,mccaughrean2004,mugrauer2006,liu2007,luhman2007,pinfield2006,pinfield2008,pinfield2012,burningham2009,burningham2013,goldman2010,albert2011,dayjones2011,luhman2011,murray2011,deacon2012a,deacon2012b,dupuy2012,wright2013}.

WISE~J200520.38+542433.9 (\W2005 hereafter) is a faint (J$\sim$19.6 mag), late-type T dwarf located 188$\farcs$5 from Wolf 1130 in the constellation Cygnus.
Wolf 1130 (LHS 482, Gl 781) is a well constrained high-proper motion system \citep{vanleeuwen2007} at $\sim$16pc from the Sun. 
Additionally, Wolf 1130 has been identified as a $\sim$12 hour period single-lined spectroscopic binary consisting of a M1.5 subdwarf and possibly a helium white dwarf \citep{gizis1998}. 
UVW space velocities in the literature identify Wolf 1130 as a member of the old disk or halo population \citep{stauffer1986,leggett1992}, 
and \citet{rojasayala2012} determine a low metallicity ([Fe/H] = $-$0.64 $\pm$ 0.17~dex) for the M1.5 subdwarf component of the system.
All of these characteristics imply an age of at least 2~Gyrs for Wolf 1130 \citep{hansen1998}, but more likely 10 - 15~Gyrs.

In this paper we present the discovery of the T8 subdwarf \W2005. 
Astrometry from space- and ground-based imaging verifies this object as having a common proper motion with Wolf 1130.
We have compiled photometry and near-infrared spectroscopy of \W2005 that show the hallmarks of low metallicity and advanced age.
In Section 2 we describe our imaging and photometry, while Section 3 describes the spectroscopic observations.
Analysis of our observations is presented in Section 4 and the primary properties of \W2005 are summarized in Section 5.

\section{Imaging \& Photometry} 
\begin{figure}
\epsscale{1}
\figurenum{1}
\plotone{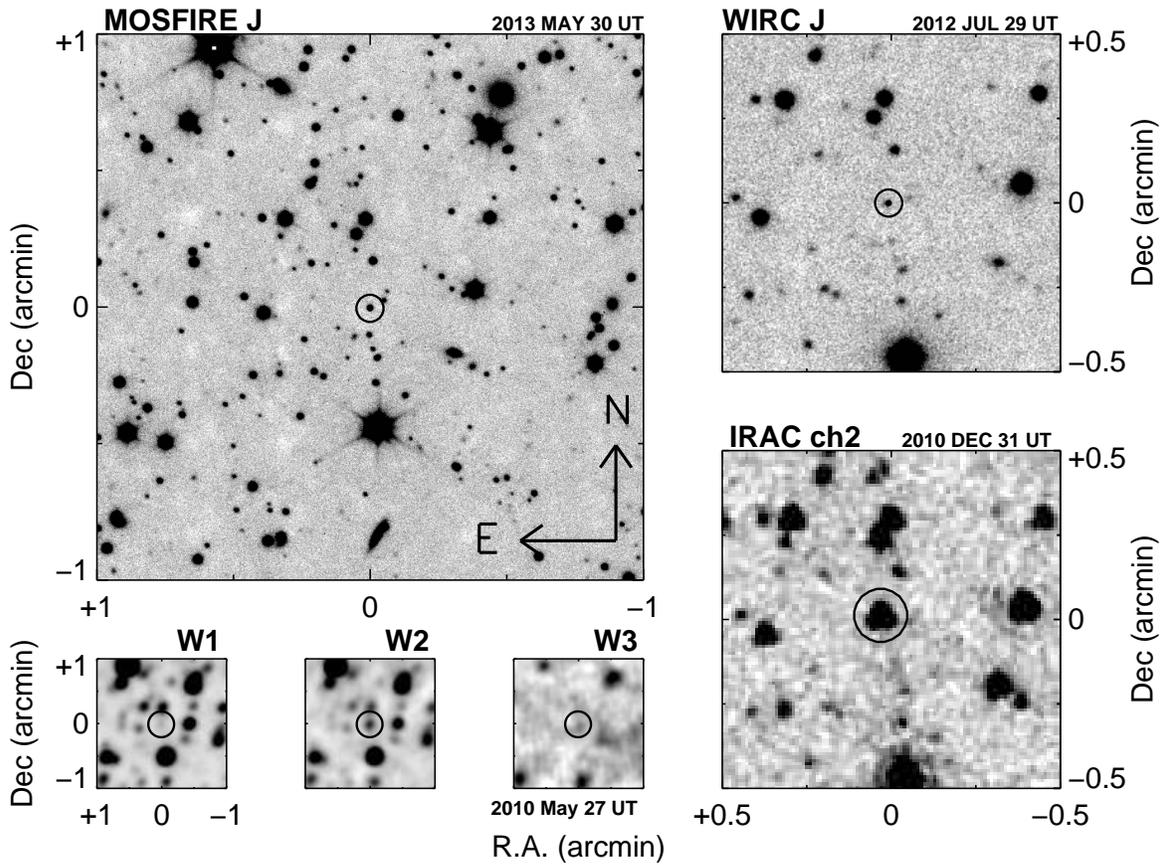}
\caption{Finder chart collection centered on \W2005. Images are labeled with the instrument, filter and UT date of observation. 
\W2005 is marked by a circle. A full list of our observations is provided in Table~\ref{astrometry}.
\label{finder}}
\end{figure}

\W2005 was selected as a brown dwarf candidate by employing the WISE All-Sky Data Release \citep{cutri2012} color selection criteria 
described by \citet{kirkpatrick2011,kirkpatrick2012} and \citet{mace2013}.
Tables~\ref{photometry} and~\ref{astrometry} list various parameters for Wolf 1130 and \W2005 from the literature and the analysis in this paper.
Figure~\ref{finder} shows a compilation of images centered on \W2005, which is too faint to be detected in the 2MASS 
survey and lies outside the UKIDSS Galactic Plane Survey area \citep{lawrence2007,lucas2008}.

\subsection{2MASS}
Photometry for Wolf 1130 is from the 2MASS All-Sky Point Source Catalog. 
In the 2MASS images Wolf 1130 is $\sim$6$\arcsec$ away from a background source.
$J$- and $H$-band measurements were unblended and derived with a single component fit. 
The $K_{s}$-band photometry for Wolf 1130 required a two component fit to deblend it from the neighboring source.

\subsection{WISE}
The {\it WISE} mission was a two-pass all-sky survey at 3.4, 4.6, 12, and 22~$\mu$m, hereafter referred to as bands W1, W2, W3, and W4, respectively \citep{wright2010}. 
\W2005 was identified by its red $W1-W2$ and $W2-W3$ colors. 
As a variant of the methane imaging technique, redder $W1-W2$ colors identify a lack of flux in 
the deep 3.3~$\mu$m CH$_4$ absorption band relative to the abundant 4.6~$\mu$m flux in T and Y dwarfs \citep{rosenthal1996,tinney2005,wright2010,mainzer2011}.

\begin{figure}
\epsscale{0.9}
\figurenum{2}
\plotone{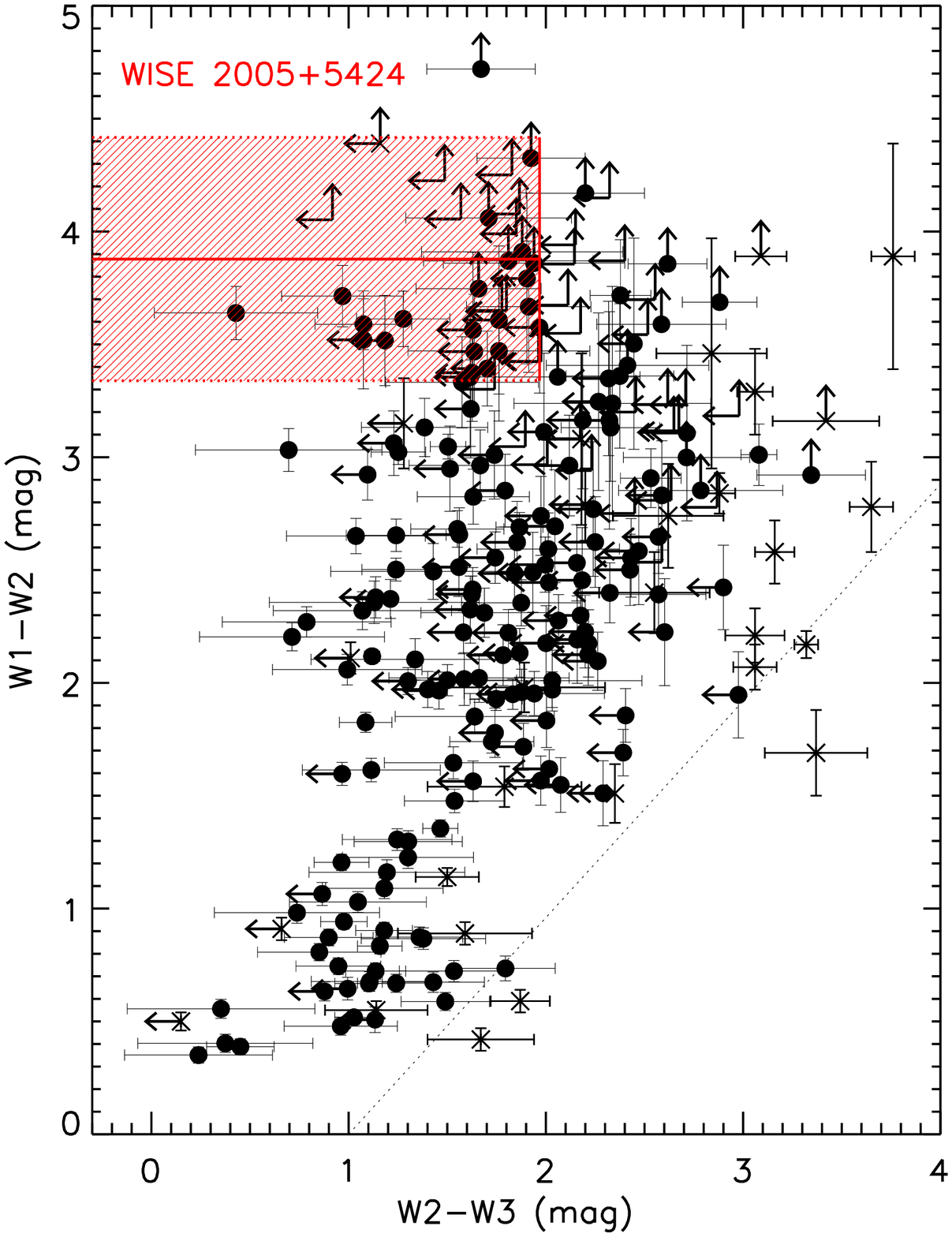}
\caption{{\it WISE} $W1-W2$ versus $W2-W3$ color-color diagram. A 1$\sigma$ error box for the \W2005 photometry is illustrated by the red shaded area. 
T dwarfs from \citet{mace2013}, \citet{kirkpatrick2011,kirkpatrick2012},  \citet{mainzer2011}, \citet{cushing2011}, and \citet{burgasser2011} 
are marked with black circles and limit arrows. Interlopers are marked by black x's and limits. 
The dotted line identifies the $W1-W2$ $>$ 0.96 ($W2-W3$) $-$ 0.96 selection criterion outlined by \citet{kirkpatrick2011}. 
\label{wise_colors}}
\end{figure}

Figure~\ref{wise_colors} shows the $W1-W2$ and $W2-W3$ colors of \W2005 along with other {\it WISE}-discovered T and Y dwarfs. 
The $W1-W2$ color of \W2005 is consistent with a type later than T6.
The $W2-W3$ color is an upper limit and is not a useful independent indicator of the object's expected spectral type, but the limit is consistent with a T or Y dwarf. 
The two {\it WISE} epochs used in our astrometric solution are from the All-Sky Data Release for the first {\it WISE} epoch 
and the average of the twenty detections in the Post-Cryo Single Exposure Source Table for the second {\it WISE} epoch.

\subsection{Spitzer}
\begin{figure}
\epsscale{0.9}
\figurenum{3}
\centering
\includegraphics[width=3.22in]{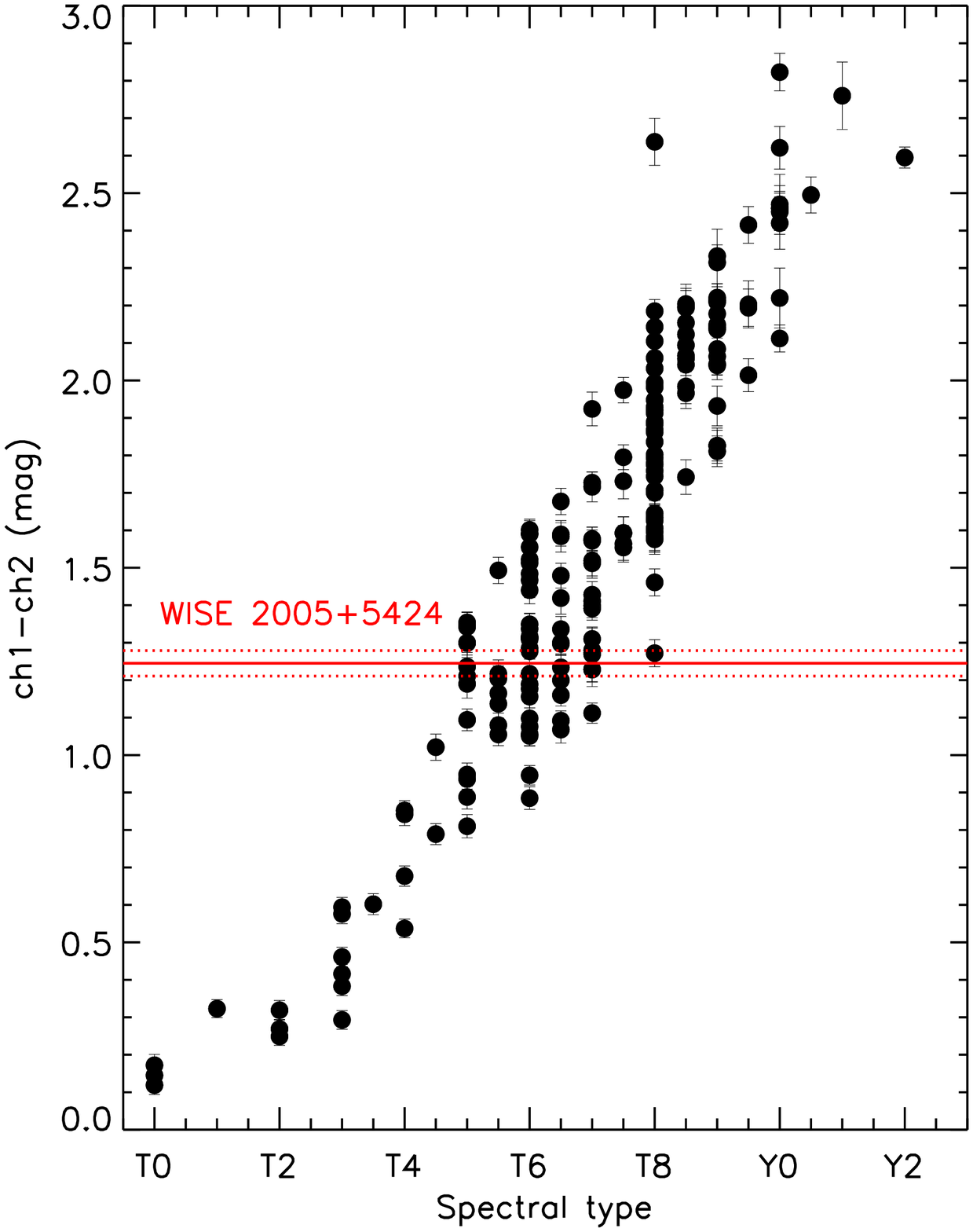}
\includegraphics[width=3.22in]{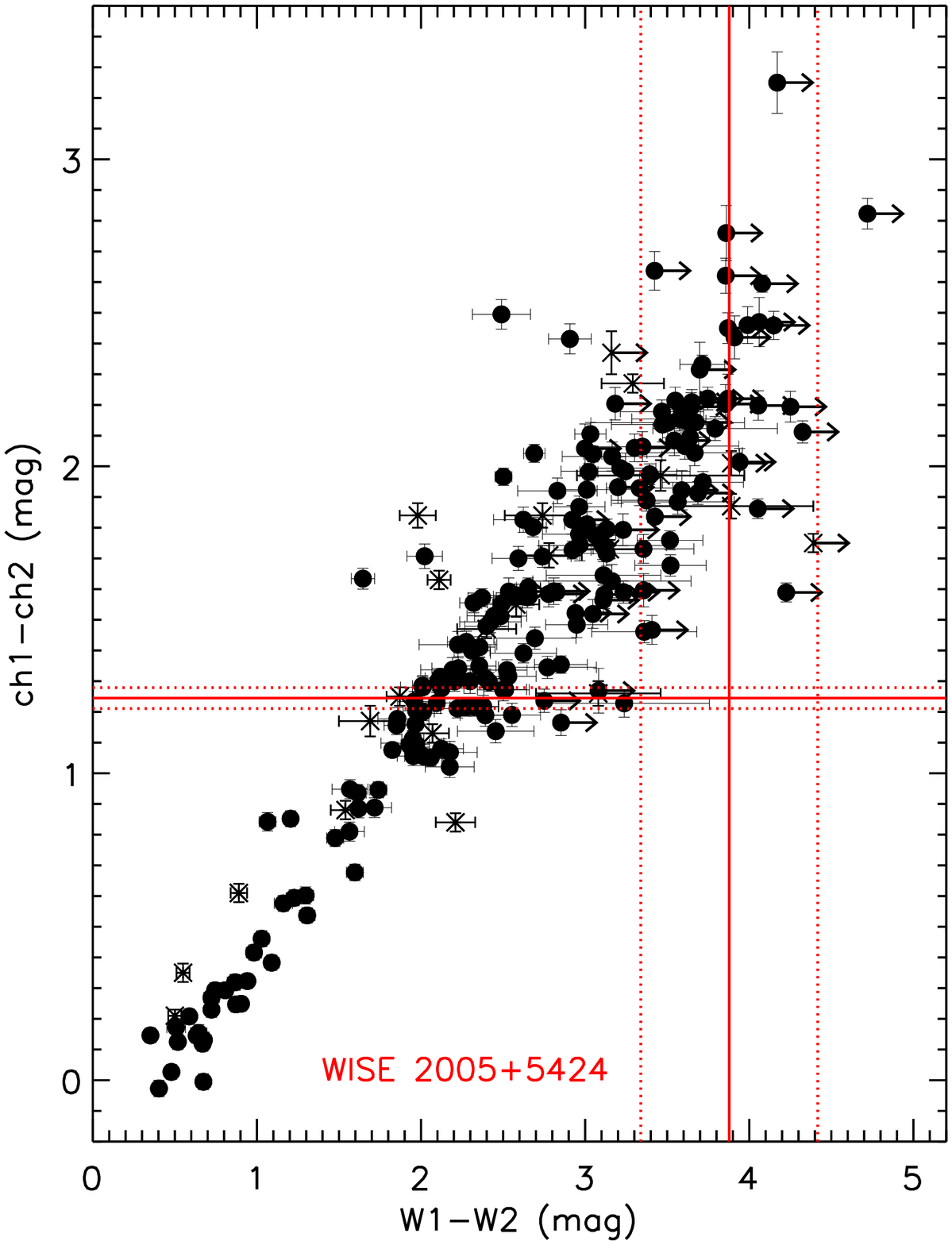}
\caption{{\it Spitzer} $ch1-ch2$ versus spectral type and {\it WISE} $W1-W2$ color. The symbols are the same as in Figure~\ref{wise_colors}.
{\it Spitzer} photometry constrains \W2005 to spectral types between T5 and T8 and \W2005 occupies a unique corner of the color-color diagram.
\label{spitzer}}
\end{figure}

The Infrared Array Camera (IRAC; \citealt{fazio2004}) onboard the {\it Spitzer Space Telescope}  (\citealt{werner2004})
was used during the warm {\it Spitzer} mission to obtain deeper, more precise 
photometry of \W2005 in the 3.6 and 4.5~$\mu$m channels (hereafter, $ch1$ and $ch2$, respectively).
Observations were made on three separate epochs as part of Cycle 7 and Cycle 8 programs 70062 and 80109 (Kirkpatrick, PI).
Data acquisition and reduction was the same as in \citet{kirkpatrick2011} and \citet{mace2013}.
The $ch1-ch2$ color as a function of spectral type is shown in Figure~\ref{spitzer}. 
This {\it Spitzer} photometry constrains \W2005 to spectral types between T5 and T8, which is a narrower range than the {\it WISE} colors imply. 
The $ch1-ch2$ versus $W1-W2$ color-color diagram is also shown in Figure~\ref{spitzer}. 
\W2005 occupies a unique corner of this color space, which we discuss in more detail in Section 4.3.

\subsection{WIRC}
\begin{figure}
\epsscale{0.9}
\plotone{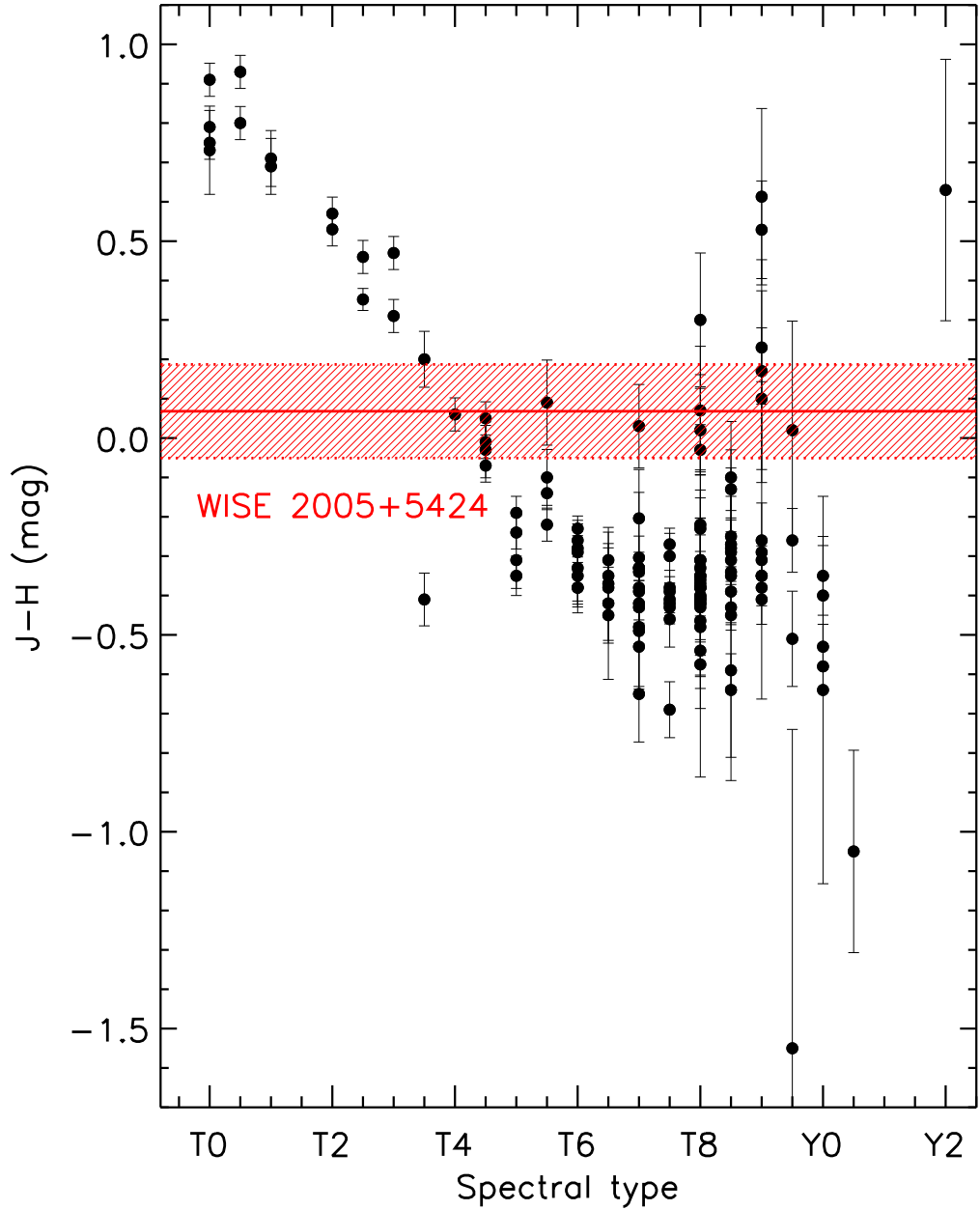}
\figurenum{4}
\caption{Palomar/WIRC ($J-H$)$_{MKO}$ color versus spectral type.  \W2005 photometry encloses the shaded area, which is redder than most late-type T dwarfs. 
Data points are collected from the literature cited in Section 2.4.
\label{JH_type}}
\end{figure}

Photometry of \W2005 on the MKO system \citep{simons2002,tokunaga2002} is from the Wide-field Infrared Camera \citep[WIRC;][]{wilson2003} at the 
5~m Hale Telescope at Palomar Observatory. Source extractions from our observations use apertures that 
are 1.5$\times$FWHM of the source point-spread function following the method outlined by \citet{kirkpatrick2011}. The MKO $J-H$ colors for a collection of objects from the literature 
\citep{leggett2010a,leggett2012,leggett2013,albert2011,kirkpatrick2011,kirkpatrick2012,dupuy2012,mace2013,burningham2010b,burningham2013,thompson2013}
are shown in Figure~\ref{JH_type}.
\W2005 is redder in $J-H$ than the T5 - T8 spectral type constrained by {\it Spitzer} and mostly consistent with late-type T dwarfs.
The red $W1-W2$ color of \W2005 results in red $J-W2$ and $H-W2$ colors, which are also consistent with a late-type T dwarf classification \citep[see Figures 7 and 8 in][]{kirkpatrick2011}.

\subsection{MOSFIRE}
The Multi-Object Spectrometer For Infra-Red Exploration \citep[MOSFIRE;][]{mclean2012} is a spectrometer and imager employed at 
the Cassegrain focus of the 10~m Keck I telescope at W.~M.~Keck Observatory. 
MOSFIRE provides imaging over a field of view of $\sim$6$\farcm$9 diameter with 0$\farcs$18 pix$^{-1}$ sampling. 
The detector is a $2K\times2K$ H2-RG HgCdTe array from Teledyne Imaging Sensors with low dark current and low noise \citep{kulas2012}.
\W2005 was observed with MOSFIRE in a box9 dither pattern with 20.37s exposures and 3 coadds, for a total exposure time of $\sim$61s per dither.
The signal-to-noise ratio for \W2005 in the final mosaic is $\sim$600.
We were able to detect \W2005 in single, sky-subtracted exposures with signal-to-noise ratios of $\sim$200, but require the full dither set to construct a quality sky flat.
The MOSFIRE filters are on neither the 2MASS nor MKO system since they are optimized for spectroscopy.
Characterization of the methane imaging technique and photometric calibration of MOSFIRE is underway (Logsdon et al., in preparation).
The astrometric coordinates derived for \W2005 are the median of the nine individual $J$-band exposures and the uncertainty is the standard deviation of the offset 
between 2MASS reference coordinates and corresponding detections in the MOSFIRE image. 

\subsection{OSIRIS}
High resolution images of \W2005 were taken on 2013 June 19~UT with the OH 
Suppressing Infra-Red Imaging Spectrograph \citep[OSIRIS;][]{larkin2006} on the
Keck I telescope.  The 
OSIRIS imager has a fixed plate scale of 0$\farcs$02 pix$^{-1}$ and a field of view of 20$\arcsec$.
We used the R = 14.1 USNO B star 1444$-$0306431 \citep{monet2003} for the input
of the Natural Guide Star \citep{wizinowich2000} wavefront sensor.  Five images were taken
in the $Hbb$ filter (a broad-band $H$ filter) with single frame exposure times of 120s. 
\W2005 has a FWHM of 0$\farcs$14 on the mosaic created from the five images.  There
are no obvious companions detected, nor does the PSF of \W2005 differ significantly
from other objects in the mosaic. 

\subsection{NIRSPEC}
Employing the slit-viewing camera (SCAM) on the Near-Infrared Spectrometer (NIRSPEC, \citealt{mclean1998,mclean2000}) at Keck Observatory,
we obtained a nodded pair of images of \W2005. 
SCAM is a 256$\times$256 HgCdTe array behind the NIRSPEC filter wheel that images a 46$\arcsec$ square field around the science slit.
Astrometric measurements from SCAM utilized ten reference stars. 
Only four of the objects in the field were detected by 2MASS and coordinates for the remaining reference stars were taken from our deeper MOSFIRE images.
Position uncertainties are estimated as the seeing (0$\farcs$45) at the time of observation.

\section{Spectroscopy}
With NIRSPEC we observed \W2005 in the medium-resolution mode with the $Y$-band (N1) and $J$-band (N3) filters. 
An $H$-band spectrum of \W2005 was obtained with MOSFIRE in spectroscopic mode.
Figure~\ref{spec} shows the $Y$-, $J$-, and $H$-band NIRSPEC and MOSFIRE spectra along with comparisons to observed and model T dwarf spectra (discussed in detail in Section 4.2).

\subsection{NIRSPEC}
On 2013 May 21 UT we observed \W2005 in the $Y$ band with the 0${\farcs}$38 slit (R$\sim$2350). 
Six 600s exposures were reduced with the publicly available REDSPEC package with modifications to remove residuals from the sky-subtracted pairs prior to 1-D spectral extraction.
The wavelength solution was derived from OH sky-lines. The A0 standard HD 199217 was observed at a similar airmass as the target for telluric correction of the spectrum. 
In the $J$ band we used the 0${\farcs}$57 slit (R$\sim$1400) on three different epochs with two different A0 standards; 2012 June 08 UT (HD 199217), 
2012 September 06 UT (HD 199217) and 2012 September 25 UT (HD 205314). 
Comparison of arclamp and sky-line locations throughout each night show no offset between the first spectrum of \W2005 and the last spectrum of the 
A0 standard and provides consistent wavelength calibrations.
The combined spectrum is a compilation of 11 nod pairs ($22\times300$s exposures) for a total integration time of 6600s. 
Final $J$-band wavelength solutions were derived from NeAr arc lamp spectra acquired before or after observations of \W2005 and the A0 standards.

\subsection{MOSFIRE}
From the {\it WISE} and NIRSPEC/SCAM coordinates we derived an initial estimate of the proper motion for \W2005.
Using this proper motion, we updated the coordinates of \W2005 and used the MOSFIRE Automatic GUI-based Mask Application 
(MAGMA)\footnote{Available at http://www2.keck.hawaii.edu/inst/mosfire/magma.html} to design a 
slitmask centered on \W2005. An $H$-band spectrum was obtained on 2012 October 12 UT with a spectral resolution of R$\sim$3500 for a 0$\farcs$7 slit width.
Observations were made using a Nod2 (AB) pattern with exposure times of 120s to minimize skyline saturation, resulting in
17 high-quality nod pairs and a total exposure time of 4080s.
For telluric correction we observed the A0 standard HD 199066 (V = 9.10) at five locations along a $15$\arcsec$\times0${\farcs}$7$ slit with 12s exposures. 
A modified version of the REDSPEC package was used for spectral extraction and telluric correction, and wavelength solutions made use of OH skylines.

\section{Analysis}

\subsection{The Companionship of \W2005 and Wolf 1130}
Wolf 1130 is a well known high-proper motion system \citep{wolf1921,ross1939,giclas1968,luyten1976,gliese1991,vanleeuwen2007} with
$\mu_{\alpha}$ = $-$1$\farcs$163 $\pm$ 0$\farcs$005 yr$^{-1}$ and $\mu_{\delta}$ = $-$0$\farcs$900 $\pm$ 0$\farcs$004 yr$^{-1}$.
Parallax measurements of 59.4 $\pm$ 2.4 mas \citep{harrington1980},  61 $\pm$ 2 mas \citep{vanaltena1995}, and 63.17 $\pm$ 3.82 mas \citep{vanleeuwen2007} place
Wolf 1130 between 14.9 and 17.5~pc from the Sun. 
It is also a single-lined spectroscopic binary \citep{joy1947,gliese1969,stauffer1986,dawson1998} with a radial velocity range of $\sim$240 km s$^{-1}$
and a center of mass velocity $\gamma \approx -$34 km s$^{-1}$ \citep{gizis1998}.
From the parameters in Table~\ref{photometry} we derive UVW velocities for Wolf 1130, relative to the Sun, of $-101\pm7$, $-44\pm2$, and $33\pm3$, respectively.
This is consistent with the old disk-halo membership requirements of \citet{leggett1992}.

Additional constraints on the age of Wolf 1130 are provided by its metallicity and binarity. 
\citet{gizis1997} presents optical spectroscopy, along with CaH and TiO spectral indices \citep{reid1995}, and derives a subdwarf M1.5 classification for Wolf 1130.
\citet{lepine2007} also classify it as a subdwarf on their expanded classification scheme. 
From a large sample of M dwarf $K$-band spectra, in which Wolf 1130 is the most iron-poor member, \citet{rojasayala2012} determine T$_{eff}$ = 3483 $\pm$ 17~K, 
[M/H] = $-$0.45 $\pm$ 0.12~dex, and [Fe/H] = $-$0.64 $\pm$ 0.17~dex.
Other [Fe/H] calculations in the literature are $-$0.80~dex \citep{woolf2009}, $-$0.87 dex \citep{stauffer1986}, $-$0.89 dex \citep{bonfils2005}, and $-$1.02 dex \citep{schlaufman2010}.
Despite the presumed old age of Wolf 1130, \citet{stauffer1986} note H$\alpha$ emission in their optical spectra, and \citet{reid1995} and \citet{gizis1998} identify 
H$\alpha$ and H$\beta$ variability. The source of this hydrogen emission may be related to the $\sim$12 hour period of Wolf 1130
derived by \citet{gizis1998} in his single-lined orbital solution.
\citet{young1987} reason that M-dwarfs in binary systems with periods less than $\sim$5 days are agitated into producing Balmer line emission as a 
result of tidally increased rotational velocities. 
Indeed, Wolf 1130 has a large rotational velocity \citep[$v$~sin~$i$ $\approx$ 15 - 30~km s$^{-1}$;][]{stauffer1986,gizis1998} relative to most other early-type 
M dwarfs \citep[$v$~sin~$i$ $\leq$ 10~km s$^{-1}$;][]{jenkins2009}.

Combining his orbital solution with the lack of secondary spectral lines, and the fact that the Wolf 1130 photometry is consistent with other M subdwarfs,
\citet{gizis1998} concludes that the companion is most likely a $\sim$0.35 M$_{\odot}$ helium white dwarf.
Based on the calculations by \citet{hansen1998} the age of Wolf 1130 is at least 2~Gyrs, but more likely 10 - 15~Gyrs.
Furthermore, as the secondary component of Wolf 1130 lost mass and transitioned into a white dwarf, common envelope evolution may have forced
the mass ratio of the system closer to unity and ensured the ejection of the envelope material \citep{ivanova2013}.
The ejection of the envelope would have reduced the separation of the sdM1.5+WD binary and produced the short period derived by \citet{gizis1998}.

A number of imaging campaigns have searched for additional companions to Wolf 1130, 
but were either not deep enough to detect the much fainter \W2005 component and/or not wide-field enough to enclose both objects.
\citet{jao2009} used the KPNO Mayall 4~m telescope and the United States Naval Observatory (USNO) speckle camera \citep{mason2006} to search for companions at 5500 $\pm$ 240 \AA\  
in only a 3$\arcsec$$\times$3$\arcsec$ field.
C.R.Gelino observed Wolf 1130 on 2005 August 01 UT with the Keck natural guide star adaptive optics in the K$_s$ filter on the NIRC2 instrument. 
Short exposures and a 40$\arcsec$$\times$40$\arcsec$ field of view excluded \W2005 and didn't reveal the white dwarf component at the 50~mas scale. 
 \citet{riaz2006} presents {\it Spitzer} observations of Wolf 1130 made between 2004 October and 2006 March and also did not detect
 \W2005 in their shallow IRAC and MIPS imaging at 3.6, 4.5, 5.8, 8, and 24~$\mu$m.

Fitting our astrometry with both the proper motion and parallax as free parameters, we derive a proper motion for \W2005 of 
$\mu_{\alpha}$= $-$1$\farcs$138 $\pm$ 0$\farcs$102 yr$^{-1}$ and $\mu_{\delta}$= $-$0$\farcs$988 $\pm$ 0$\farcs$106 yr$^{-1}$, 
which is the same as the published values for Wolf 1130 to within 1$\sigma$.
Our position measurements are not precise enough to strongly constrain the parallax of \W2005 and we derive an insignificant parallax of 0$\farcs$019 $\pm$ 0$\farcs$096.
The astrometric measurements for \W2005 are compiled in Table~\ref{astrometry}. 
Figure~\ref{cpmfit} illustrates the agreeable match in our astrometric fits when proper motion and parallax are left as free parameters (dashed lines) or forced to be the same as Wolf 1130 (solid lines). 
The fit with parallax and proper motion as free parameters gives
$\chi^2 = 6.72$ for 13 degrees of freedom, while the fit forced to match
the motion of Wolf 1130 gives $\chi^2 = 8.07$ for 16 degrees of freedom.
Based on the companionship criteria of \citet[][$\Delta$$X$ = (($\mu$/0.15)$^{-3.8}$$\Delta$$\theta$$\Delta$$\mu$)$^{1/2}$ $<$ 1]{LB2007} and \citet[][$\Delta$$\mu$/$\mu$ $<$ 0.2]{dupuy2012} 
it is unlikely that \W2005 and Wolf 1130 ($\Delta$$X$ $\approx$ 0.07 and $\Delta$$\mu$/$\mu$ $\approx$ 0.12) are a chance alignment.
Assuming that the distance to \W2005 is the same as Wolf 1130 (15.83 $\pm$ 0.96~pc), from the current angular separation (188$\farcs$5) we calculate a projected separation of $2985 \pm 181$ AU.
The orbital period at this distance is on the order of 200,000 years.
As discussed by \citet[][Section 3.1]{dayjones2011}, a brown dwarf in a system slowly losing mass would migrate outward by to up to $\sim$4 times its original distance.
Before this migration \W2005 would have been as close as 700~AU from Wolf 1130.

For a distance of 15.83 $\pm$ 0.96~pc, we derive the absolute $H$ and $W2$ magnitudes that are given in Table~\ref{photometry}.
These values are shown in Figure~\ref{absmags} along with the late-type T dwarfs and magnitude-spectral type relations from \citet{kirkpatrick2012}. 
The absolute magnitudes we determine for \W2005 are consistent with late-type T dwarfs, but slightly fainter than the T8 spectral type we assign in Section 4.2.
The low luminosity of \W2005 for its spectral type implies that it is not a close, equal mass binary with another brown dwarf since binaries are generally located above the absolute 
magnitude versus spectral type relations \citep{looper2008}.
Our Keck/OSIRIS $H$-band imaging, described in Section 2.6, does not reveal a companion to \W2005 at 0$\farcs$14 ($\sim$2.2 AU).
Of course, we can not fully rule out the possibility of a fainter Y subdwarf companion \citep{leggett2013}.
The faintness of \W2005 also implies that \W2005 has a smaller radius and is older than other T8 dwarfs. 

\citet{burrows2003} model solivagant T dwarfs with temperatures between $\sim$130 and 800~K with masses from 1 to 25~M$_J$ and solar metallicities to show that the $J-H$, $J-W2$, and $H-W2$
colors redden as late-type T dwarfs age. This reddening is also seen in other models \citep{saumon2008, morley2012} 
and in the observed colors of late-type T dwarfs \citep{kirkpatrick2012,leggett2013}. 
\citet{burrows2006} also consider non-solar metallicities for temperatures between 700 and 2200~K.
The 3 - 6~$\mu$m flux of the 700~K models predict the reddest $J-W2$ and $H-W2$ colors for older objects 
with high surface gravities and low metallicities. 
\citet{stephens2009} constrain the effective temperature of a T8 to between 600 and 900~K.
\W2005 displays the reddened colors associated with a late-type T subdwarf, which agrees with the age of Wolf 1130 of at least 2~Gyrs (and likely 10 - 15 Gyrs).
\citet{baraffe2003} predict that at 1~Gyr, a brown dwarf in our assumed range of T$_{eff}$ would have a mass of $\sim$0.020~M$_{\odot}$ and radius 0.100 R$_{\odot}$. 
At 10~Gyrs, a brown dwarf of this temperature would have a mass $\sim$0.050~M$_{\odot}$ and radius 0.079~R$_{\odot}$.
\citet{burrows2011} model brown dwarf radii and find that old, low-metallicity brown dwarfs are $\sim$10 - 25$\%$ smaller than younger, solar metallicity brown dwarfs.
If \W2005 is old and metal-poor then it is likely small. This decreased size partially explains why the absolute magnitudes are fainter than we expect for the T8 spectral type. 

\subsection{Comparing \W2005 to Observed and Synthetic Spectra}
\begin{figure}
\epsscale{1.0}
\figurenum{5}
%\centering
\includegraphics[width=2.1in]{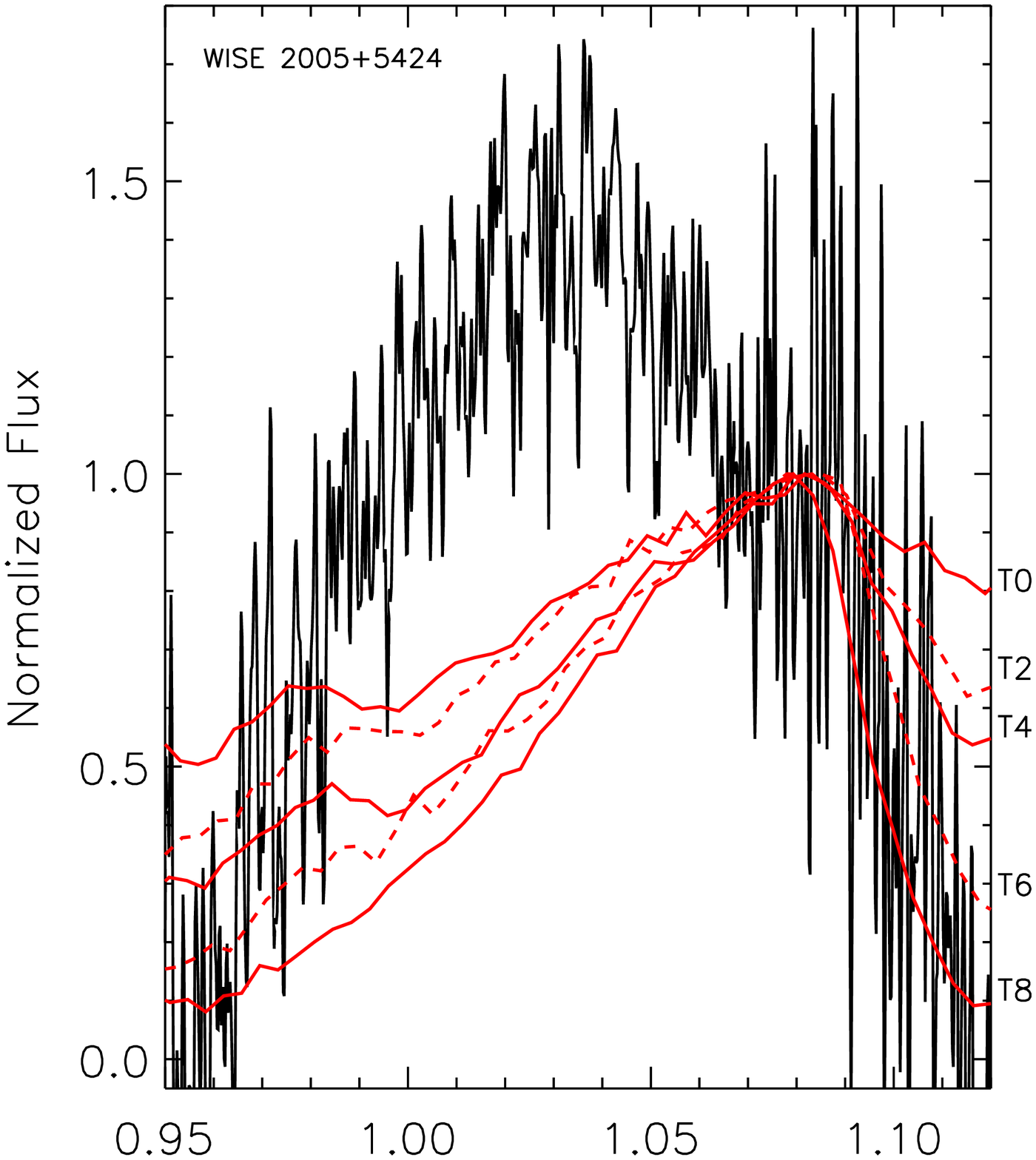}
\includegraphics[width=2.1in]{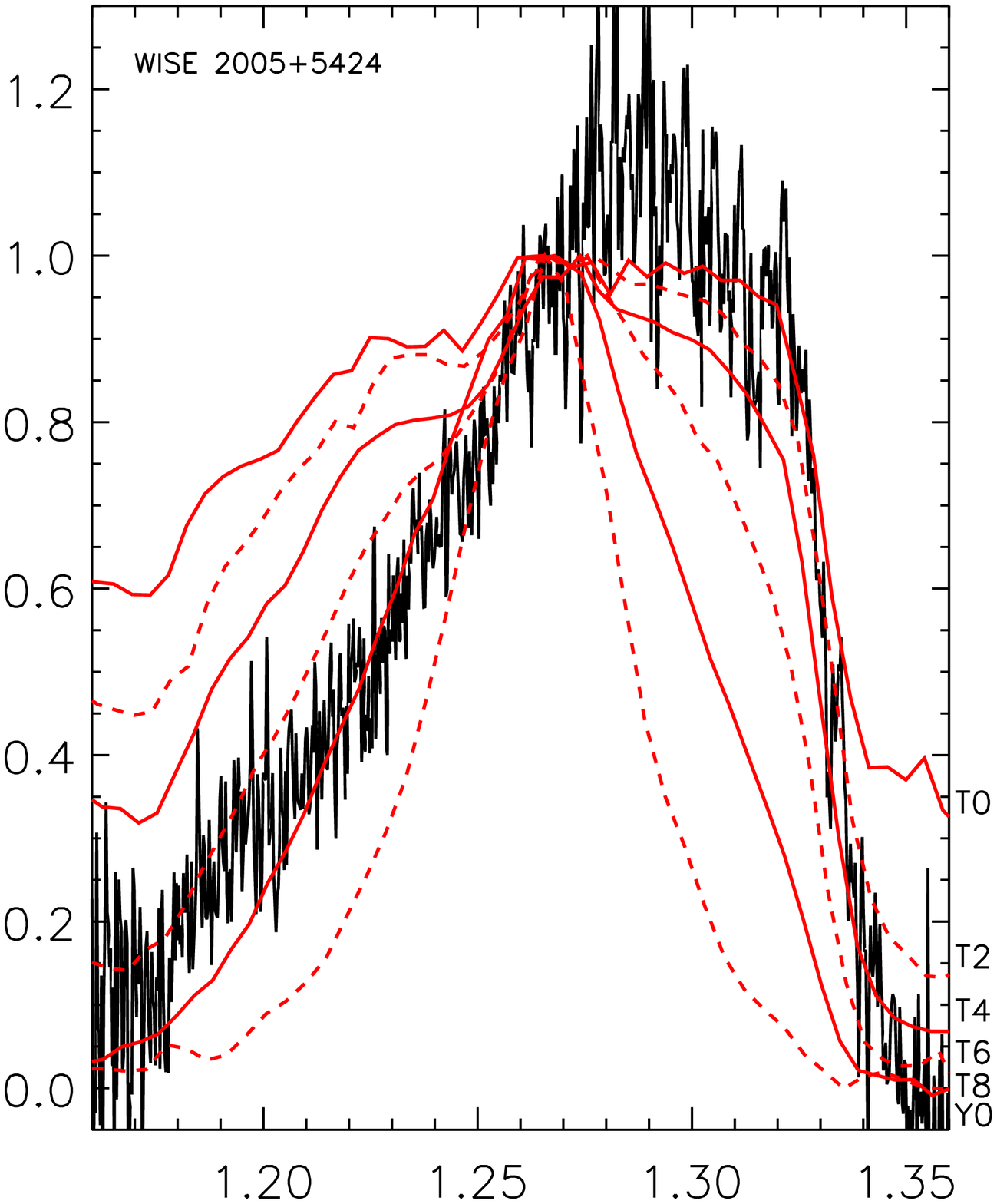}
\includegraphics[width=2.1in]{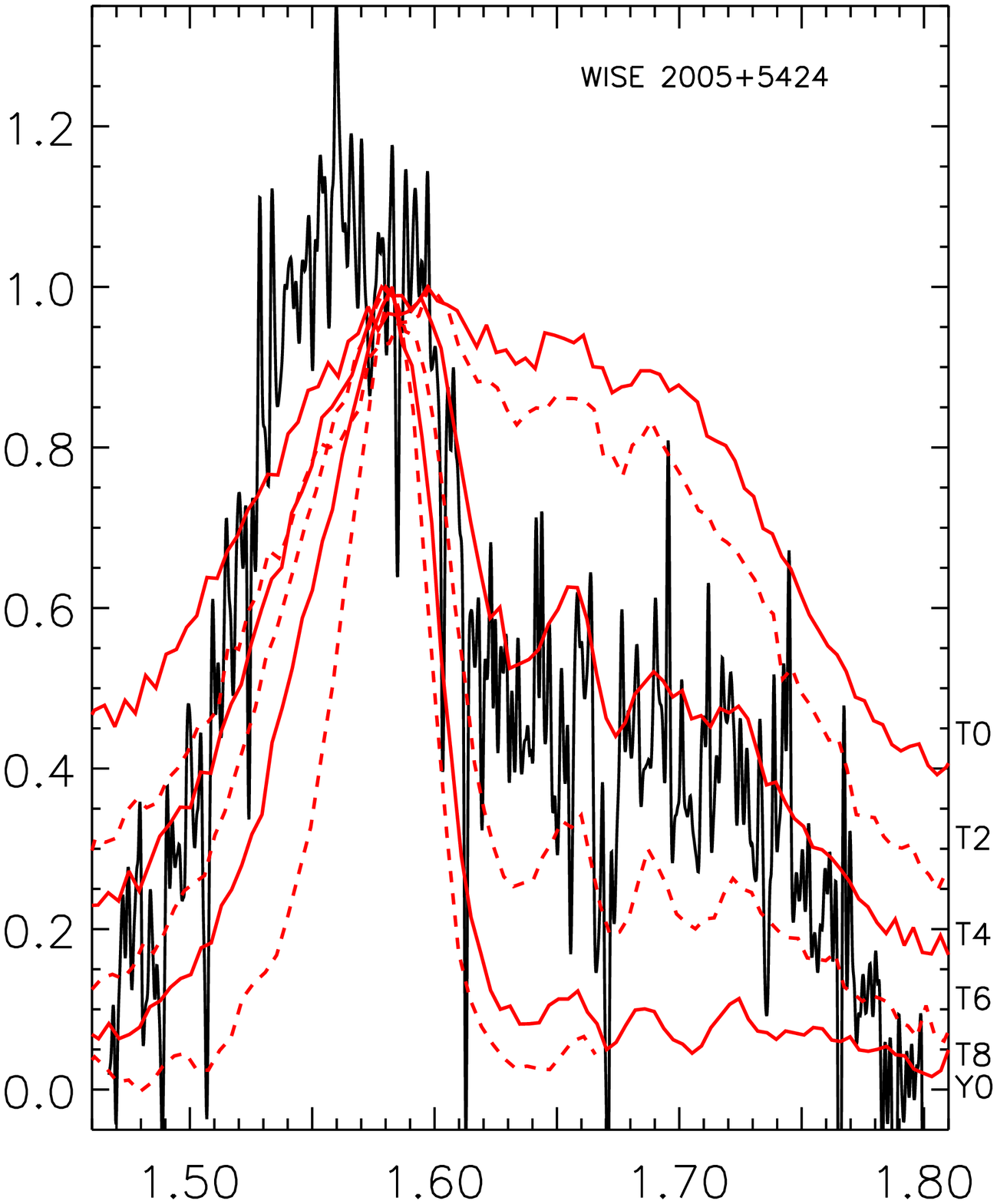}
\includegraphics[width=2.1in]{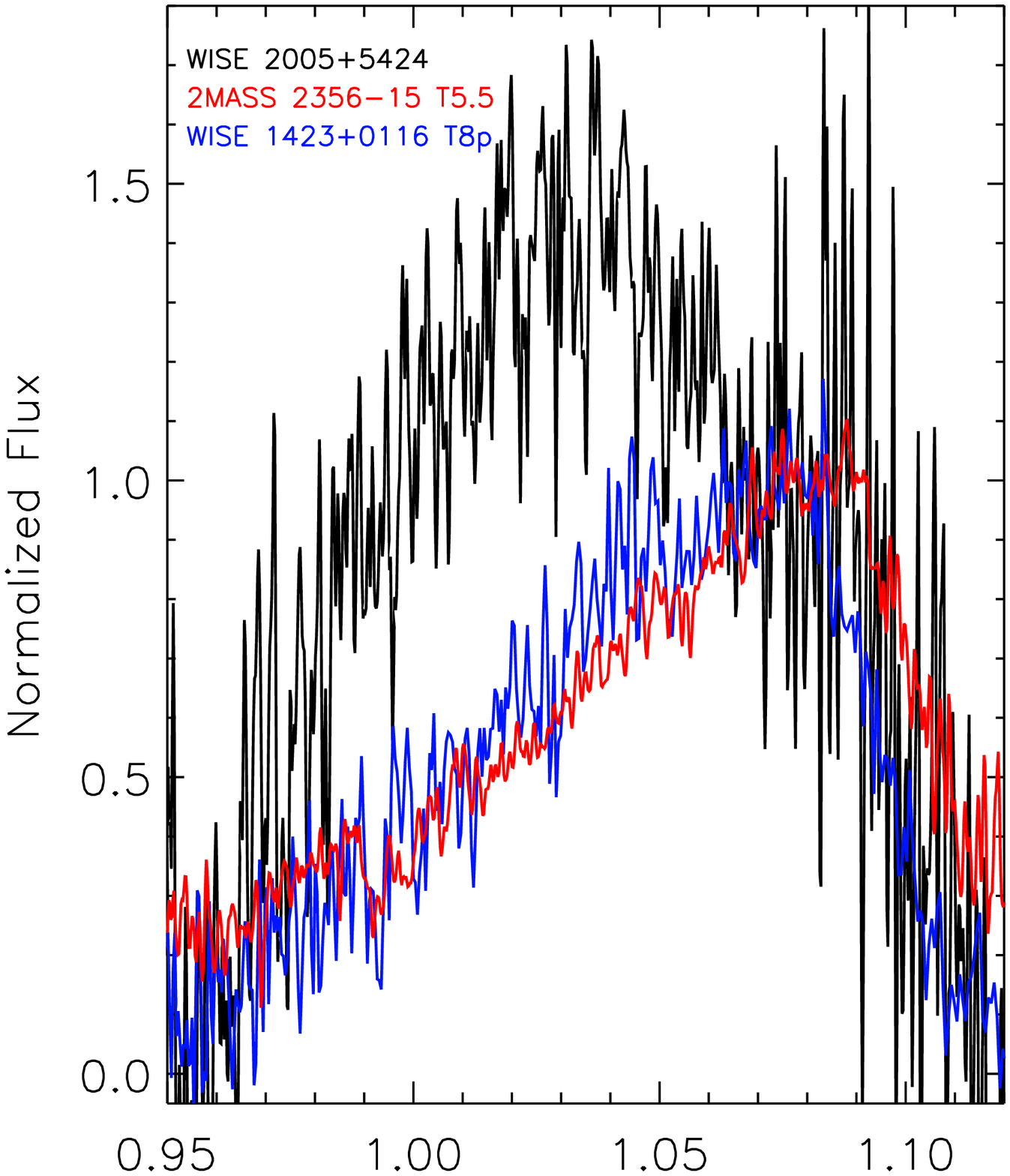}
\includegraphics[width=2.1in]{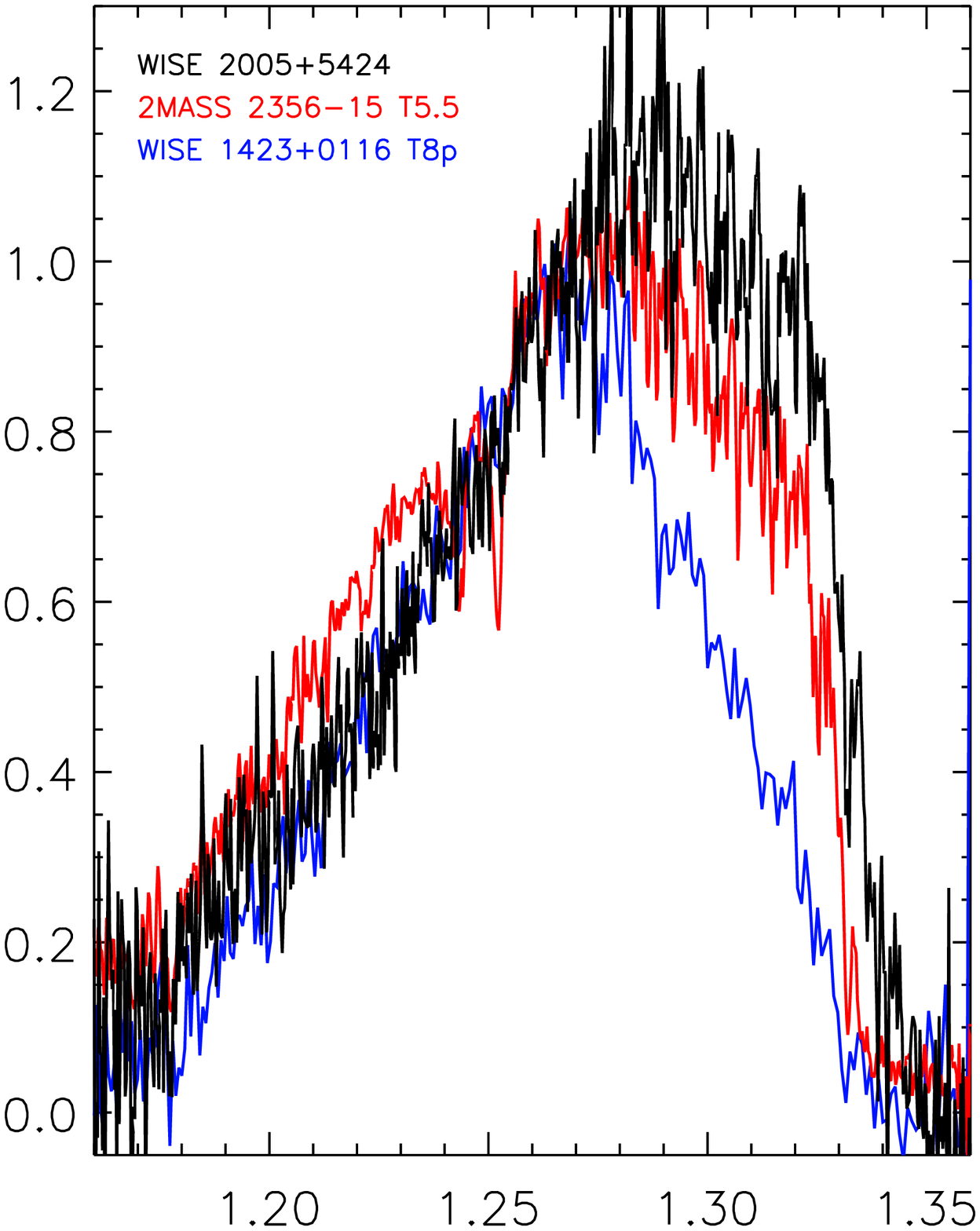}
\includegraphics[width=2.1in]{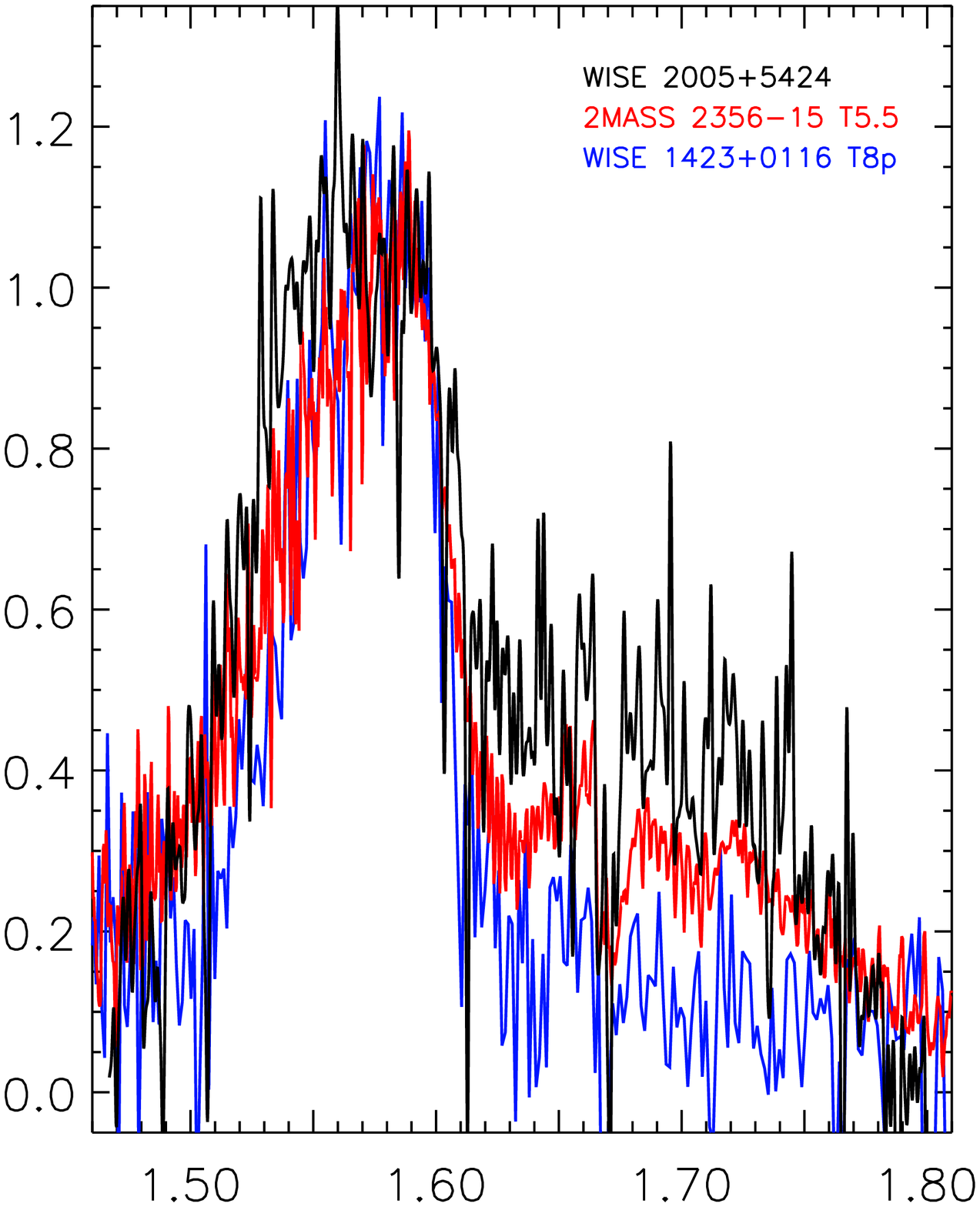}
\includegraphics[width=2.1in]{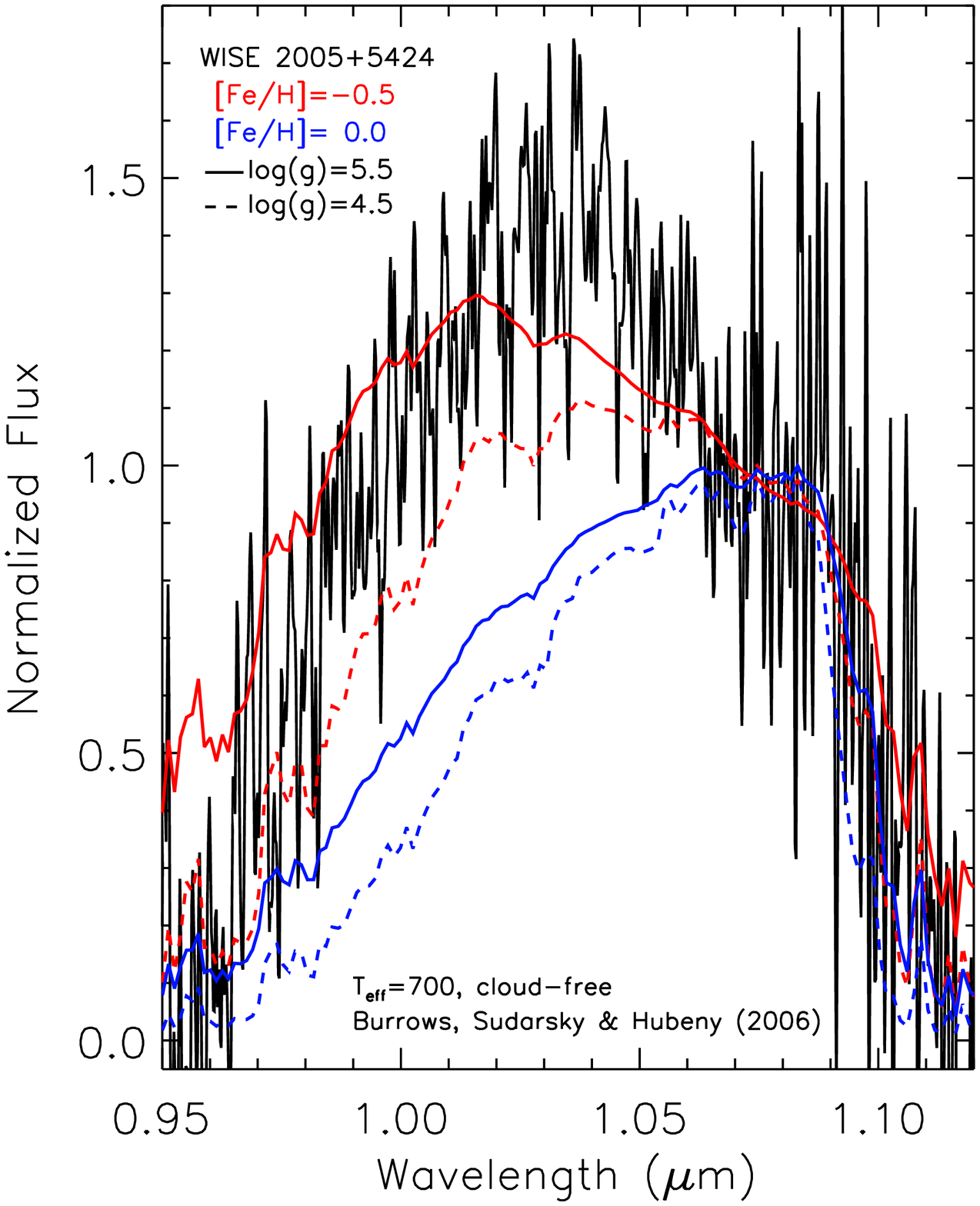}
\includegraphics[width=2.1in]{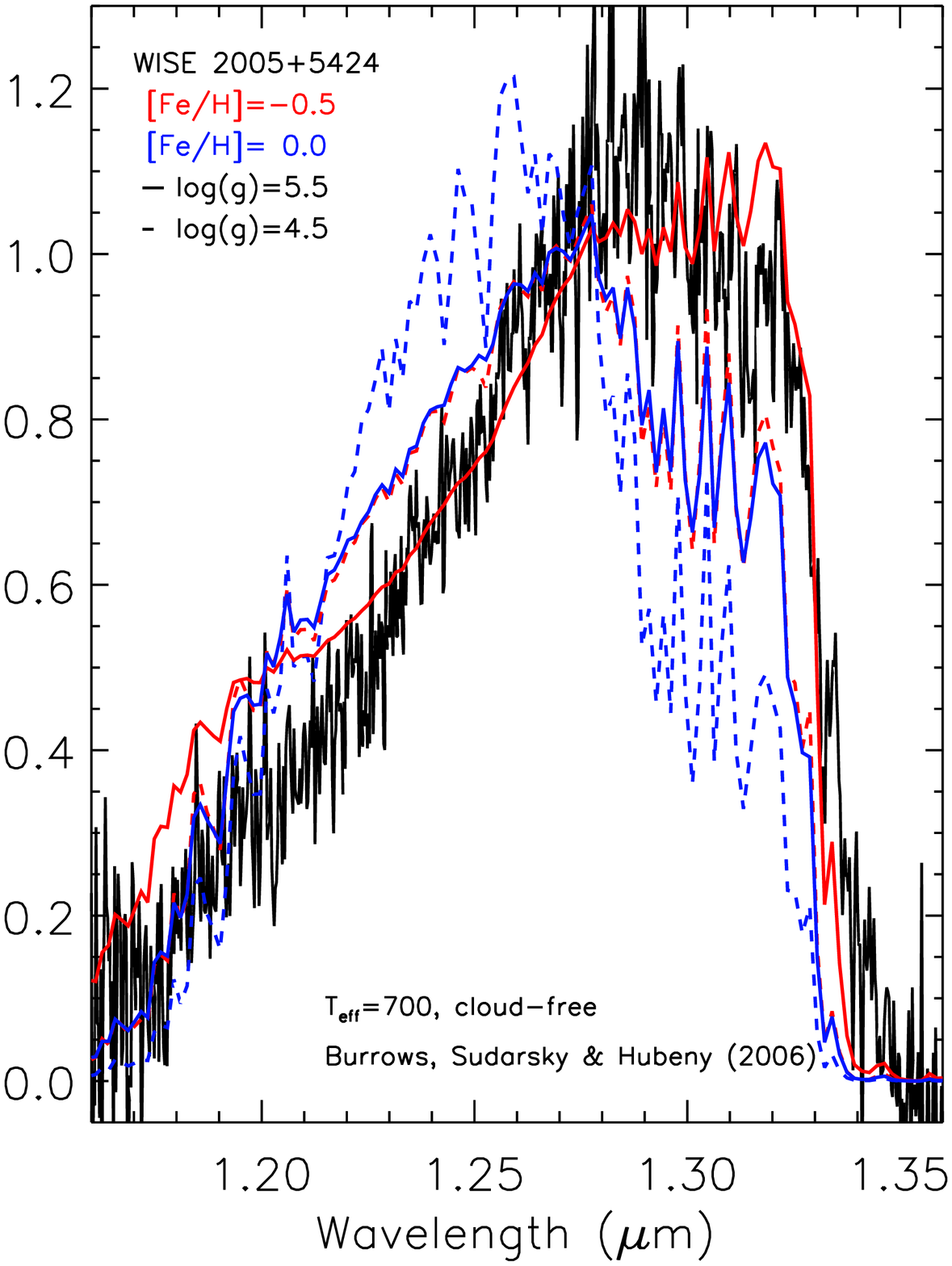}
\includegraphics[width=2.1in]{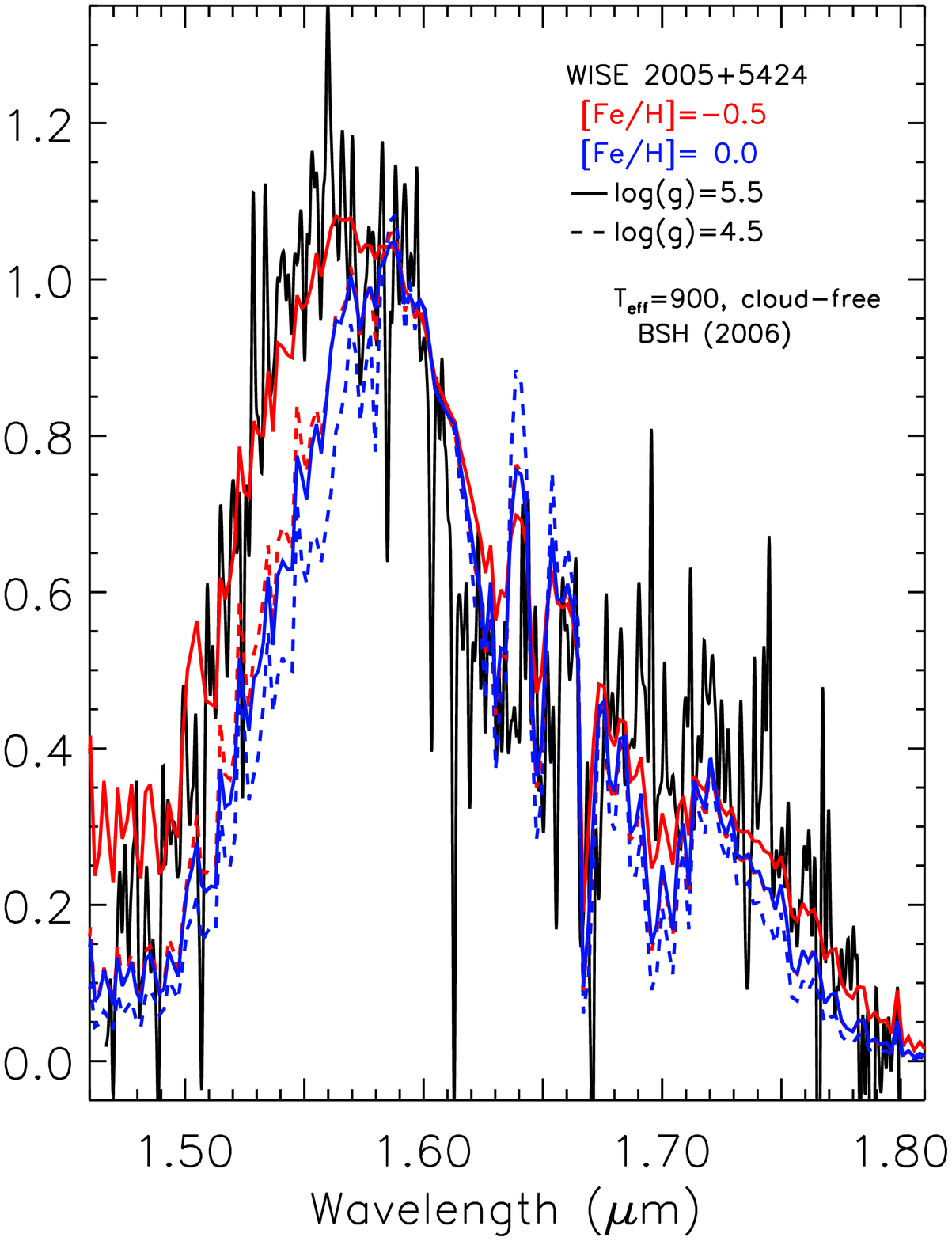}
\caption{NIRSPEC $Y$- and $J$-band and MOSFIRE $H$-band spectra of \W2005 (solid black lines) normalized at $\sim$1.08, 1.28, and 1.59~$\mu$m respectively.
For comparison, the top row includes the T0, T2, T4, T6, T8, and Y0 spectral standards \citep[red lines, labeled on the axis;][]{burgasser2006a,cushing2011}.
The middle panels include the T5.5 dwarf 2MASS J23565477$-$1553111 (solid red line) and peculiar T8 dwarf WISE J142320.84+011638.0 (solid blue line).
In the bottom row we include the best matching cloud-free models from \citet{burrows2006} for the various log(g), [Fe/H], and T$_{eff}$ values listed.
\label{spec}}
\end{figure}

\begin{figure}
\epsscale{0.9}
\figurenum{6}
\plotone{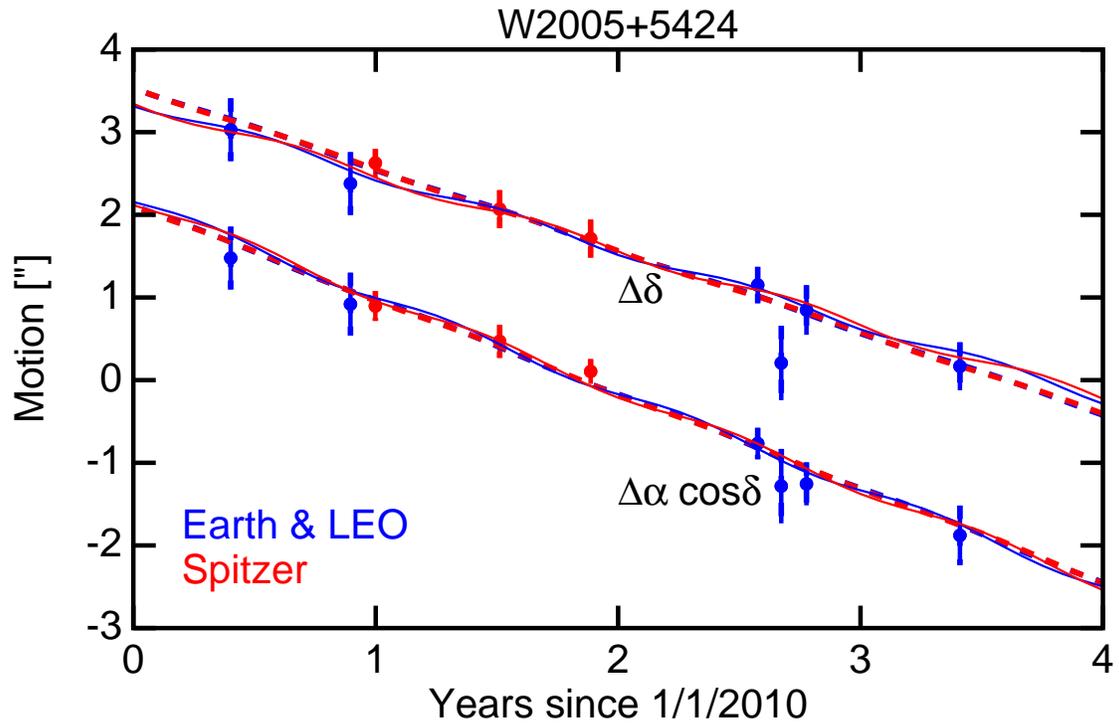}
\caption{Astrometric fits for \W2005. The blue curves and points are for ground-based or Low Earth Orbit observatories, while the red curves and points are for {\it Spitzer}. 
The dashed lines show the fit with proper motion and parallax as free parameters, while the lighter solid curves show the fit forced to match Wolf 1130.
$\Delta\delta$ has been displaced by a constant for clarity.
\label{cpmfit}}
\end{figure}

\begin{figure}
\epsscale{0.9}
\figurenum{7}
\centering
\includegraphics[width=3in]{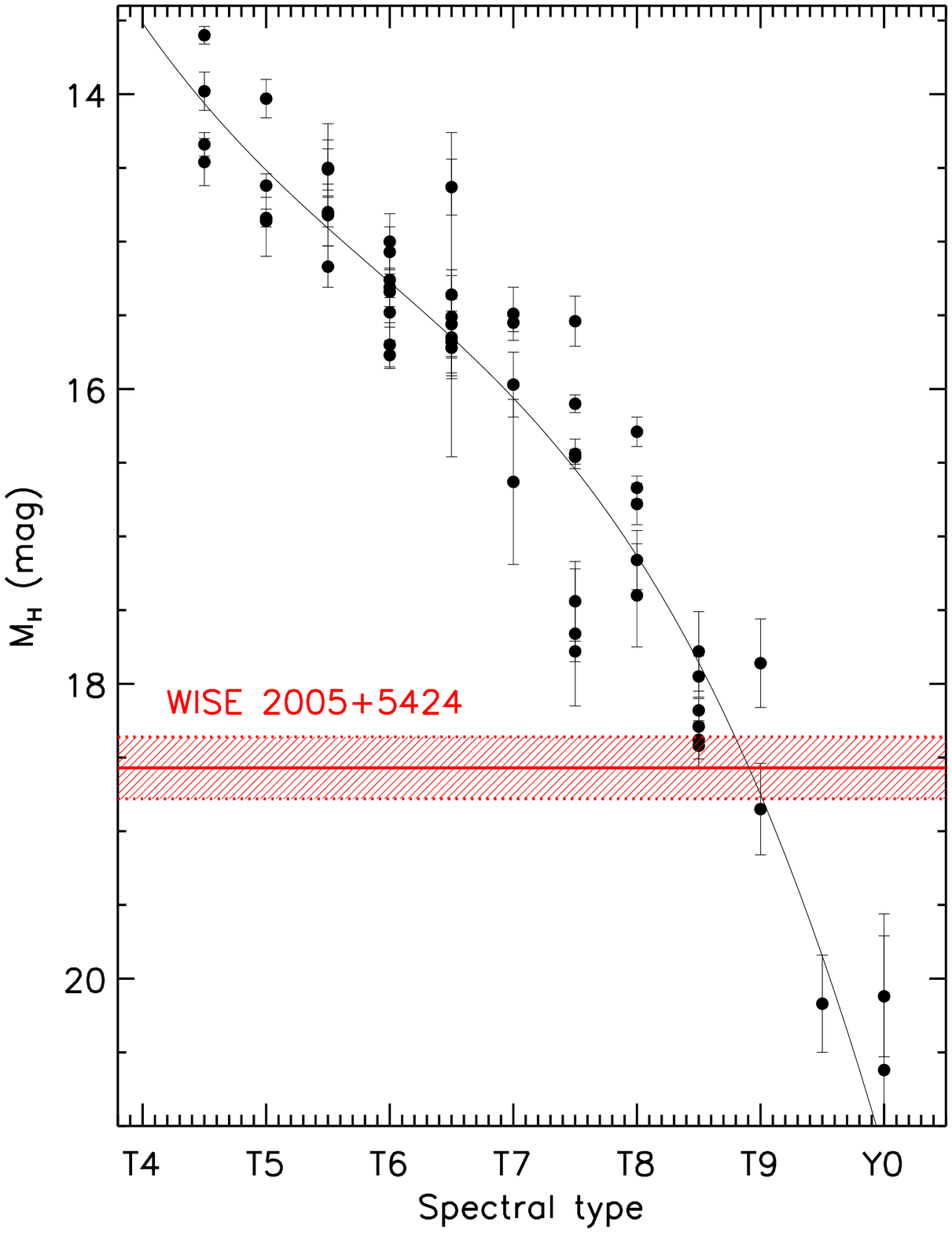}
\includegraphics[width=3in]{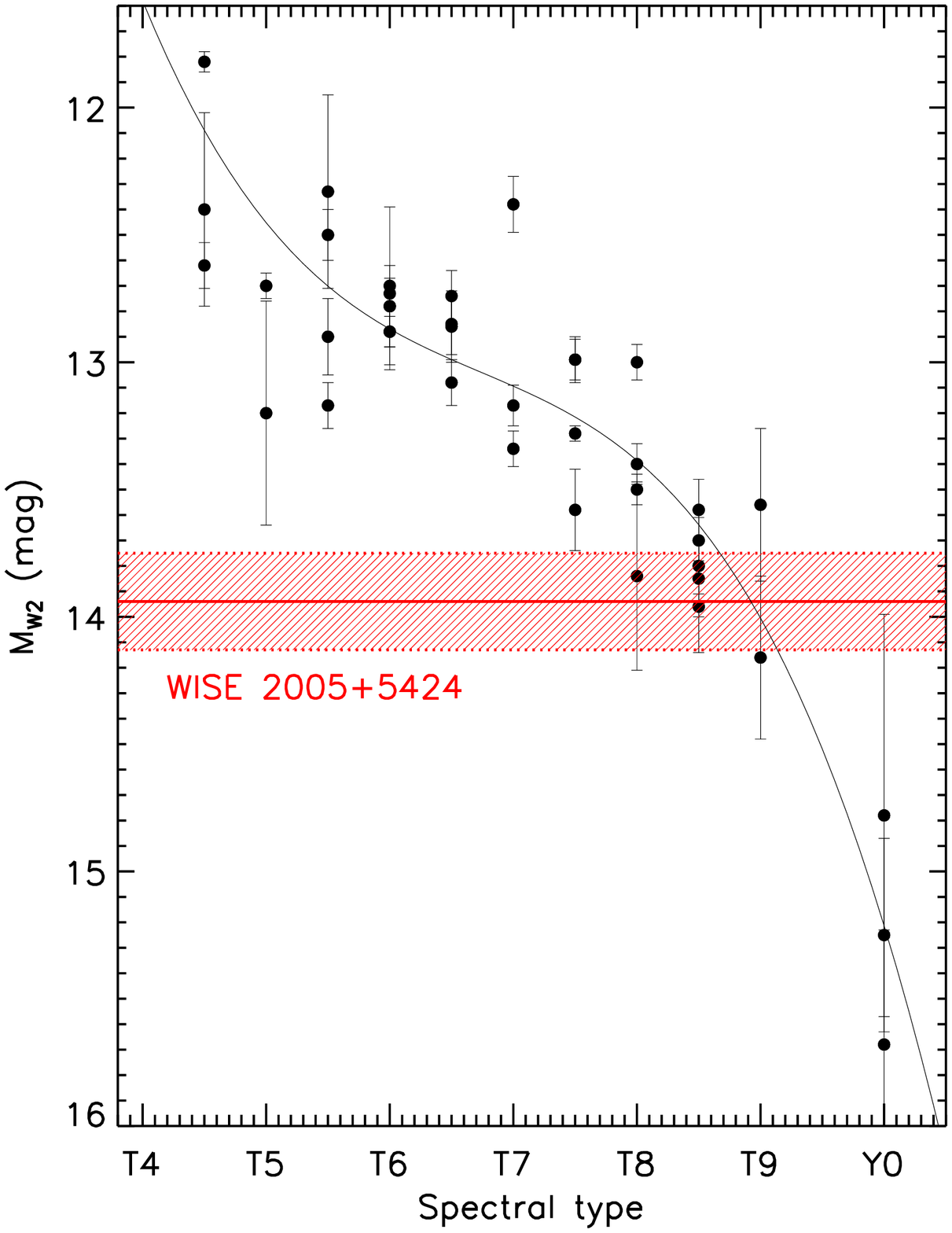}
\caption{Absolute $H$ and $W2$ magnitudes versus spectral type \citep[Figures 12 and 13; ][]{kirkpatrick2012}. The range for WISE 2005+5424 (assuming a distance of 15.83~pc) is shown
by the shaded region. Both absolute magnitude ranges are below the trend lines at the T8 spectral type that we assign, which supports its subdwarf classification.
\label{absmags}}
\end{figure}

In Figure~\ref{spec} we present $Y$-, $J$-, and $H$-band spectroscopy of \W2005.
Without significant overlap between the individual bands, and because of the low signal-to-noise of the spectra, we are not able to reliably flux calibrate the bands relative to each other.
To produce a meaningful comparison we normalize each of our spectra at the standard flux peak for that band; $Y$ ($\sim$1.08~$\mu$m), $J$ ($\sim$1.28~$\mu$m), and $H$ ($\sim$1.59~$\mu$m).
The top panels of the figure show the T0, T2, T4, T6 and T8 spectral standards of 
\citet{burgasser2006a}\footnote{Spectra from the SpeX Prism Spectral Library; maintained by Adam Burgasser} and the Y0 standard 
(WISEP~J173835.52+273258.9) defined by \citet{cushing2011}.
The middle row compares our observed spectra to the T5.5 dwarf 2MASS~J23565477$-$1553111\footnote{Spectra from the BDSS; \citet{mclean2003,mclean2007}; http://bdssarchive.org/} \citep{burgasser2002,burgasser2006a} 
and the peculiar T8 dwarf WISE~J142320.84+011638.0 \citep[BD +01\degr 2920B;][]{pinfield2012, mace2013}.
The 2MASS 2356$-$1553 spectra were published by \citet{mclean2003} and represent the mid-type T dwarf classification that some of our photometry imply.
WISE 1423+0116 is the closest object of similar type to \W2005 in the literature and the Magellan/FIRE spectrum from \citet{mace2013} is used here for comparison.
Discovered by \citet{pinfield2012}, WISE 1423+0116 is a companion to a G1 dwarf at $\sim$17.2~pc, which provides an inferred [Fe/H] = $-$0.36 $\pm$ 0.06~dex.
Based on comparison to the model isochrones of \citet{baraffe2003} they estimate T$_{eff}$ = 680 $\pm$ 55 K and log(g) = 5.0 $\pm$ 0.3~dex for WISE 1423+0116.
The bottom panels of Figure~\ref{spec} present the cloud-free models from \citet{burrows2006} with [Fe/H] = 0 and $-$0.5~dex and log(g) = 4.5 and 5.5~dex. 

Our $Y$-band spectrum of \W2005 shows the strongest hallmarks of low-metallicity. 
Comparison with the spectral standards reveals that the peak $Y$-band flux of \W2005 has
shifted from the standard peak location of $\sim$1.08~$\mu$m to $\sim$1.03~$\mu$m. 
As depicted in Figure 3 of \citet{burgasser2006BBK}, this shift and brightening at bluer wavelengths is a well predicted trait of low-metallicity T dwarfs.
No other T dwarf in the literature shows this broad morphology, which makes a $Y$-band spectral type difficult to determine based on the spectral standards alone. 
The only feature that is consistent with the standards is the red wing of the $Y$ band, which is most like a T8.
Comparing the observed templates to \W2005 we also see that WISE 1423+0116 is slightly brighter than 2MASS 2356$-$1553 between 1.00 and 1.06~$\mu$m, 
but much like \W2005 in the red wing.
\citet{burrows2006} provide models only down to 700~K, which gives a good match to \W2005. 
It is possible that lower temperature models may provide an even better match.
In any case, the 700~K solar metallicity models shown in Figure~\ref{spec} are similar to the observed templates, while the $-$0.5 dex metallicity models best match \W2005.

Comparison of the \W2005 $J$-band spectrum to the spectral standards and observed templates again emphasizes its uniqueness.
The width of the $J$-band flux peak does not follow the smooth transition that defines the spectral sequence.
The blue wing of the $J$ band is distorted relative to the other T dwarfs, while the red wing contains the bulk of the flux.
A similar $J$-band morphology was seen by \citet{burningham2010a} and \citet{burgasser2010} in the low-metallicity, high-gravity T7.5 dwarf 
SDSS J141624.08+134826.7B (although with different strengths, implying slight spectral variability of the feature on the timescale of about a month).
Additionally, the KI doublet at 1.243 and 1.254~$\mu$m is seen in the T5.5 comparison spectrum but is absent for \W2005.
This is a distinct indicator of its low-temperature and/or high surface gravity \citep{mclean2003}. 
We do not decide on a spectral type for the $J$-band spectrum of \W2005 and therefore give a range between T5 and T8.
The same 700~K models from \citet{burrows2006} demonstrate degeneracy between metallicity and gravity in the overlap between the low-gravity, low-metallicity (red dashed) case 
and the high-gravity, solar-metallicity (blue solid) one. It is only at the extremes that the $J$ band can differentiate between the different model parameters.
Just as in the $Y$ band, \W2005 is bracketed by the [Fe/H] = $-$0.5~dex models with log(g) = 4.5 - 5.5~dex at a temperature of 700~K. 
Higher temperature models produce wider $J$ band fluxes and less agreement with \W2005, but lower temperature models may provide a better match. 

In the $H$ band, \W2005 shows fewer inconsistencies with the observed templates, but its continuum is most like a mid-type T dwarf, rather than the late-type that the other bands imply.
The best matching \citet{burrows2006} models have an effective temperature of 900~K and show the same gravity-metallicity degeneracies as the $J$ band.
However, \W2005 is well-matched to the [Fe/H] = $-$0.5~dex and log(g) = 5.5~dex model. 

From our model comparisons, we estimate the spectral type of \W2005 as T8, with an effective temperature between 600 and 900~K.
This is consistent with what \citet{stephens2009} derives for a T8, but is at the limits of that study. 
Although the $H$-band spectrum is better matched with higher temperature models and an earlier spectral type, 
these characteristics would be completely inconsistent with the other spectroscopic bands and with much of our photometry.
\citet{burningham2008} identify ULAS J101721.40+011817.9 as a peculiar T8 that is similar to \W2005 since it is more like a T6 in the $H$ and $K$ bands.
Through comparison to the BT-Settl models \citep{allard2003}, but without a $Y$-band spectrum to constrain metallicity well, they estimate a large surface gravity 
(log(g) = 5 - 5.5~dex) and a wide range of ages between 1.6 and 15~Gyrs. However, none of the solar metallicity models they use provide an especially good fit to the spectrum. 

\subsection{Identifying T Dwarfs with Color Indices Similar to \W2005}
The location of \W2005 in {\it WISE} color space is unusual, but not unique.
The standout objects in Figure~\ref{wise_colors} have $W2-W3$ $\leq$ 1.3 and are WISE J000517.48+373720.5 (T9), WISE J045853.89+643452.5 (T8.5), 
WISE J062309.94$-$045624.6 (T8), WISE J115013.85+630241.5 (T8), WISE J121756.90+162640.8 (T9), and WISE J152305.10+312537.6 (T6.5p) \citep{kirkpatrick2011,mace2013}.
This subset includes WISE 0458+6434, a T8.5+T9 binary \citep{mainzer2011,gelino2011,burgasser2012}, and WISE 1523+3125, a peculiar T6.5 dwarf that \citet{mace2013} identifies as having
the same low-metallicity hallmarks as 2MASS J09373487+2931409 \citep[T6pec; ][]{burgasser2002}.

Figure~\ref{spitzer} demonstrates that in {\it Spitzer} and {\it WISE} colors \W2005 is only coincident with 
WISE J174556.65+645933.8 (T7), WISE J190903.16$-$520433.5 (T5.5), WISE J223729.52$-$061434.4 (T5), and
WISE J233543.79+422255.2 (T7) \citep{kirkpatrick2011,mace2013}. 
However, none of these objects are noted as peculiar in the limited near-IR photometry and spectroscopy currently compiled for them.
The primary difference between the {\it WISE} and {\it Spitzer} passbands, as they pertain to late-type brown dwarfs \citep[Figure 2; ][]{mainzer2011}, is that the cutoff 
for the {\it Spitzer} $ch2$ band is at $\sim$5~$\mu$m and the $W2$ cutoff is closer to $\sim$5.2~$\mu$m.
Approximately 10 - 15$\%$ of the emergent flux is in this $\Delta$0.2~$\mu$m window, which produces a slightly redder $W1-W2$ color relative to $ch1-ch2$ \citep{leggett2013}
regardless of the impacts of non-equilibrium chemistry \citep{saumon2007,hubeny2007} and sulfide or iron/silicate clouds at wavelengths just below $\sim$5~$\mu$m \citep{morley2012}.

\begin{figure}
\epsscale{0.9}
\figurenum{8}
\plotone{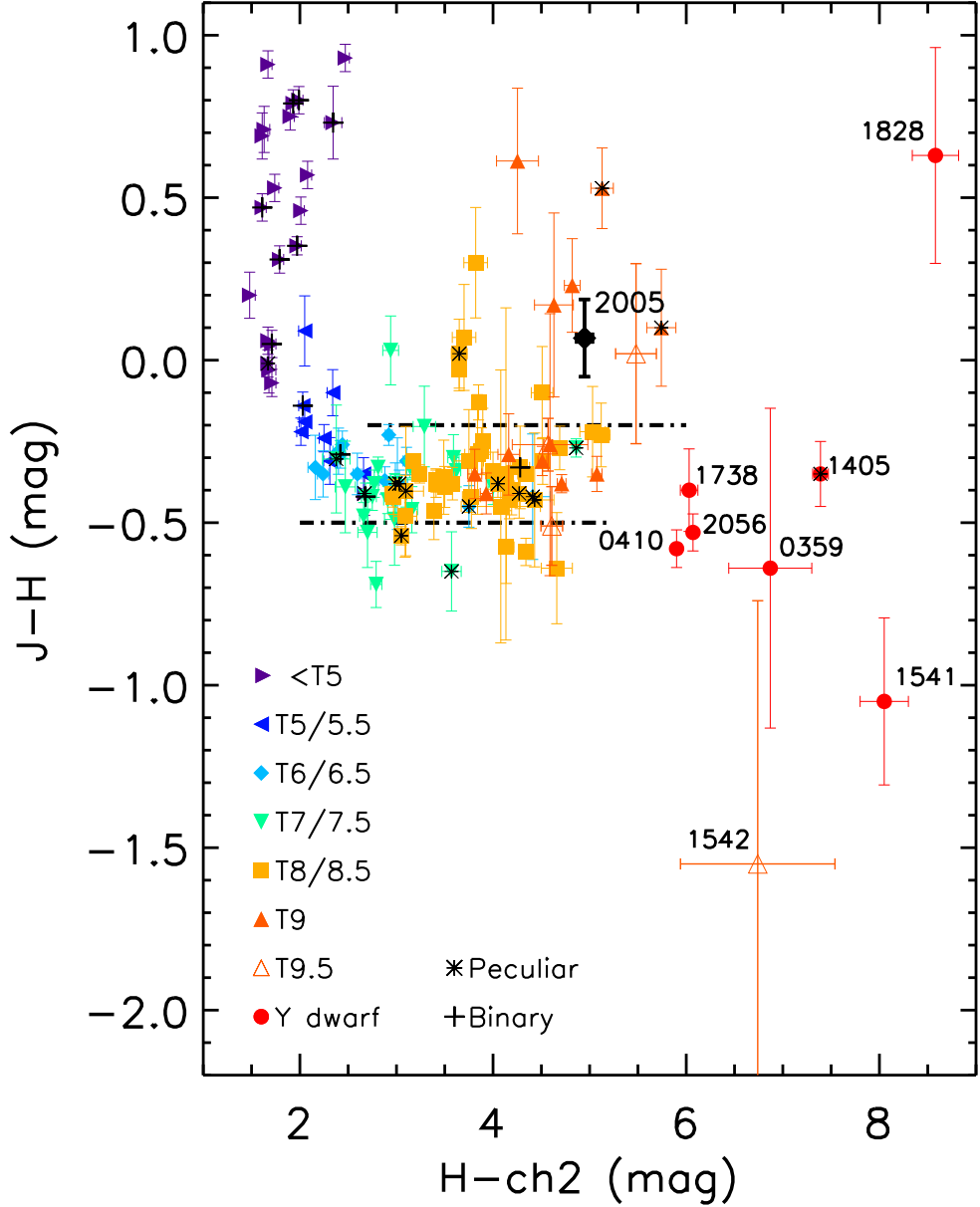}
\caption{$J-H$ versus $H-ch2$ color-color diagram. Data points are on the MKO filter system and collected from the literature referenced in Section 2.4. 
Various colored symbols differentiate the spectral type bins.
Peculiar or binary objects in the literature are marked additionally. We have marked by name WISE 2005+5424 (this paper), WISE 1542+2230 \citep[T9.5;][]{mace2013}, and the 
Y dwarfs WISE 0359$-$5401 (Y0), WISE 0410+1502 (Y0), WISE 1405+5534 (Y0p), WISE 1541$-$2250 (Y0.5), WISE 1738+2732 (Y0), WISE 1828+2650 ($\geq$Y2), 
and WISE 2056+1459 (Y0) with photometry from \citet{cushing2011}, \citet{kirkpatrick2012}, and \citet{leggett2013}. Old, metal-poor late-type T dwarfs like \W2005 have 
$J-H$ $> -0.2$, while young, metal-rich late-type T dwarfs have $J-H$ $< -0.5$ (dash-dotted lines).
\label{JH_Hch2}}
\end{figure}

Recently, \citet{burningham2013} presented a new sample of color and proper-motion selected T dwarfs from the UKIDSS Large Area Survey. 
Figure 12 of that work presents the $J-H$ versus $H-ch2$ colors of T dwarf benchmarks and the compilation of T dwarf colors from \citet{leggett2010a}.
We expand upon that plot in Figure~\ref{JH_Hch2} of this paper, which shows the $J-H$ versus $H-ch2$ color-space for T and Y dwarfs,
by including additional photometry for objects in Figure~\ref{JH_type}.
This expanded sample indicates that there is a well defined sequence for the bulk of the population, but that there are also distant outliers.
The general trend toward bluer $J-H$ colors as a result of significant methane absorption in the $H$ band is a well known feature of the T dwarf sequence.
At later spectral types, which corresponds to redder $H-ch2$ colors and lower effective temperatures, the dispersion does increase and the late-type T dwarfs begin to turn slightly redward.
With the exception of WISE J154214.00+223005.2 and WISE J182831.08+265037.7 the Y dwarfs are distinctly redder than the T dwarfs in $H-ch2$ and bluer in $J-H$.

WISE 1542+2230 was discovered by \citet{mace2013} and classified as a T9.5 from a Hubble Space Telescope Wide Field Camera 3 \citep{kimble2008} spectrum. 
The median flux spectral indices are mostly consistent with a T9.5 but the $J$-narrow index is closer to a Y0 than a T9.5 \citep{mace2013}.
Additional photometry of this source would improve its placement in Figure~\ref{JH_Hch2} and reveal if it is actually bluer in $J-H$ than most Y dwarfs.
WISE 1828+2650 was discovered by \citet{cushing2011} and initially classified, based on its equal-height $J$- and $H$-band peaks, as a $>$Y0 dwarf.
\citet{kirkpatrick2012} further expanded the Y dwarf population and with the introduction of the Y1 spectral type moved WISE 1828+2650 to a later type of $\geq$Y2.
WISE 1828+2650 is still a distant outlier in Figure~\ref{JH_Hch2}. Its $H-ch2$ color is consistent with the latest Y dwarf classification that it has been given, 
but its $J-H$ color is only matched by the largest T dwarf outliers.

Table~\ref{JHoutliers} lists the T dwarfs later than T6 that have $J-H$ colors greater than $-$0.2 and less than $-$0.5.
With $J-H$ = 0.068 $\pm$ 0.119, \W2005 is adjacent to 13 other red outliers that are primarily from {\it WISE} \citep{kirkpatrick2011,mace2013,thompson2013}. 
The exception is the T8 dwarf ULASJ095047.28+011734.3, which was first presented by \citet{leggett2012}.
\citet{burningham2013} identify ULAS J0950+0117 as a companion to LHS 6176, which is an M4 dwarf with [Fe/H] = $-$0.30 $\pm$ 0.1.
The red outliers discovered with {\it WISE} were classified from Keck/NIRSPEC $J$-band spectra or Magellan/FIRE near-IR spectra.
In the cases where $J$-band spectra were used for classification, identification of peculiar objects is not possible without broader wavelength coverage.
As discussed in \citet[][Section 5 and Appendix]{mace2013}, a number of the late-type T dwarfs have a larger $Y/J$ index and a smaller $K/J$ 
index than the spectral standards defined by \citet{burgasser2006a} and \citet{cushing2011}.
Although these objects all have bluer $Y-J$ and $J-K$ colors (inferred from the $Y/J$ and $K/J$ indices) they were not classified as peculiar since they were all observed with 
the same instrument (Magellan/FIRE) and the possibility of an instrumental bias was too strong to ignore.
The addition of $J-H$ colors in identifying these objects as outliers implies that the objects listed in Table~\ref{JHoutliers} should also be classified as peculiar, and that there is not an 
inherent bias in the FIRE observations or reductions from \citet{mace2013}.

On the opposite side of the $J-H$ color sequence there are nine blue outliers, which are listed in Table~\ref{JHoutliers}.
\citet{albert2011} discuss the peculiar T7 dwarf CFBDS J030135.11$-$161418.0 as an extremely red ($H-K_s$ = 0.92 $\pm$ 0.12) outlier in Figure 7 of that paper.
CFBDS J0301$-$1614 is also a red outlier in their $J-K_s$ color and a blue outlier in $J-H$. 
Through comparisons to BT-Settl models, \citet{albert2011} identify CFBDS J0301$-$1614 as exhibiting either low gravity (log(g) = 3.5 - 4.0) or high-metallicity.
These are both signs of extreme youth.
WISE J161705.74+180714.1 and WISE J181210.85+272144.3 were presented by \citet{burgasser2011} as low-gravity, late-type T dwarfs, with WISE 1617+1807
matching cool ($\sim$600~K) and cloudy \citet{saumon2008} models while WISE 1812+2721 looks nearly identical to Wolf 940B.
Wolf 940B is estimated as an object of intermediate gravity and age \citep{leggett2010b}. However, \citet{burgasser2011} derive a different
mass and age for WISE 1812+2721 that is very similar to WISE 1617+1807.
These differences underscore the difficulties and inconsistencies in constraining solivagant brown dwarf properties with atmospheric models alone.
The enigmatic T8 dwarf ULAS J101721.40+011817.9 (discussed in Section 4.2) is also in this group of blue $J-H$ outliers, despite having some similar features to \W2005.

From the general properties of the $J-H$ outliers it is tempting to say that T dwarfs with $J-H \leq -$0.5 are young and low-mass.
Also, $J-H$ colors greater than $-$0.2 appear to correlate with high-gravity, low-metallicity and old age.
Yet, there are exceptions to each of these categories and the influence of binarity, clouds, and non-equilibrium chemistry could also produce these distant outliers.
For example, \citet{burningham2013} report $J-H$ = $-$0.38 $\pm$ 0.04 for ULAS J0950+0117, which is below the red-outlier requirement of $-$0.2,
while Palomar/WIRC photometry from \citet{mace2013} provides the 0.020 $\pm$ 0.106 color that we use to make our selection.
Additionally, the UKIDSS photometry listed in \citet{mace2013} produces a color between these two values of $J-H$ = $-$0.2 $\pm$ 0.15.
However, it is unclear if the variability in these colors is real and only photometric monitoring can provide complete verification.

There are also young and old benchmarks that fall within the bulk $J-H$ color sequence and are not selected by the criteria that we outline.
These include the young T8.5 dwarf Ross~458C \citep[$J-H$ = $-$0.36 $\pm$ 0.03;][]{goldman2010, burningham2011} and the old T8 dwarf
WISE~J142320.84+011638.0 \citep[$J-H$ = $-$0.41 $\pm$ 0.08;][]{mace2013, pinfield2012} that we use for spectral comparison in Section 4.2.
The addition of \W2005, which has an inferred metallically of [Fe/H] = $-$0.64 $\pm$ 0.17, to the group of red $J-H$ outliers makes a compelling 
argument that this group at least partially represents the low-metallically T dwarf population.
Additionally, the position of WISE 1828+2650 in Figure~\ref{JH_Hch2} argues that its distance from the other 
Y dwarfs may be a symptom of a characteristically low-metallicity in addition to a low temperature (see Section 6.6 of \citet{burningham2013} for additional discussion).

\section{Conclusions}
The unique spectral features of WISE~J200520.38+542433.9 make spectral classification against the standards difficult, but photometric colors and 
600 to 900~K models with log(g) = 5.0 - 5.5~dex and [Fe/H] = $-$0.5~dex are in general agreement with near-infrared spectra.
Photometry and spectroscopy of \W2005 support its classification as a T8 subdwarf with common proper motion to the single-lined spectroscopic sdM1.5+WD binary Wolf 1130.
Based on this companionship, \W2005 has an inferred distance of $15.83\pm0.96$~pc and a metallicity of [Fe/H] = $-$0.64 $\pm$ 0.17.
Hallmarks of high-gravity and low-metallicity are identified in our photometric and spectroscopic analysis.
As the sample of late-type brown dwarf benchmarks is expanded to include more extreme temperatures, ages, metallicities and gravities 
we can develop improved comprehensive models.
It is only through direct comparison of observational data with these well constrained models that we can fully understand the bulk of the solivagant population.

\section{Acknowledgments}

This publication makes use of data products from the {\it Wide-field Infrared Survey Explorer}, which is a 
joint project of the University of California (UC), Los Angeles, and the Jet Propulsion Laboratory (JPL) / California 
Institute of Technology (Caltech), funded by the National Aeronautics and Space Administration.
We thank the Infrared Processing and Analysis Center (IRAC) at Caltech for funds provided by the Visiting Graduate Fellowship for G.N.M.
This publication also makes use of data products from 2MASS. 2MASS
is a joint project of the University of Massachusetts and IPAC/Caltech, funded by the National Aeronautics and Space Administration 
and the National Science Foundation (NASA). 
This research has made use of the NASA/IPAC Infrared Science Archive (IRSA),
which is operated by JPL, Caltech, under contract with NASA. Our research has benefited from the M, L, and
T dwarf compendium housed at DwarfArchives.org, whose server was funded by a NASA Small Research Grant, 
administered by the American Astronomical Society.
The Brown Dwarf Spectroscopic Survey (BDSS) is hosted by UCLA and provided an essential comparison library for our moderate-resolution spectroscopy.
This research has benefited from the SpeX Prism Spectral Libraries, maintained by Adam Burgasser at {http://pono.ucsd.edu/$\sim$adam/browndwarfs/spexprism}.
We are also indebted to the SIMBAD database, operated at CDS, Strasbourg, France. 
This work is based in part on observations made with the {\it Spitzer Space Telescope}, which is
operated by JPL, Caltech, under a contract with NASA. Support for this work was provided by NASA through an award issued to program 70062 by JPL/Caltech. This work
is also based in part on observations made with the NASA/ESA {\it Hubble Space Telescope}, obtained
at the Space Telescope Science Institute (STScI), which is operated by the Association of Universities for
Research in Astronomy, Inc., under NASA contract NAS 5-26555. These observations are associated with 
program 12330. Support for program 12330 was provided by NASA through a grant from the STScI.
The Keck/OSIRIS observations were supported by a NASA Keck PI Data Award, administered by the NASA Exoplanet Science Institute.
The spectroscopic data presented herein were obtained at 
the Keck Observatory, which is operated as a scientific partnership among Caltech, UC and NASA. 
The Observatory was made possible by the generous financial support of the W.M. Keck Foundation.
In acknowledgement of our observing time at Keck we further wish to recognize the very significant 
cultural role and reverence that the summit of Mauna Kea has always had within the indigenous Hawai'ian 
community. We are most fortunate to have the opportunity to conduct observations from this mountain.  
We thank the anonymous referee for detailed and thoughtful recommendations to improve this paper prior to publication.

\clearpage
																											
\begin{deluxetable}{lrrl}						
\tabletypesize{\small}	
\tablenum{1}
\tablewidth{4in}						
\tablecaption{Properties of Wolf 1130 and \W2005 \label{photometry}}
\tablehead{						
\colhead{} &						
\colhead{Wolf 1130} &						
\colhead{WISE 2005+5424}						
}						
\startdata							
$\alpha_{{\it WISE}}$ (J2000) &	20h 05m 00.82s	&	20h 05m 20.38s		\\
$\delta_{{\it WISE}}$ (J2000) &	+54d 25m 53.9s	&	+54d 24m 33.9s		\\
$l_{{\it WISE}}$	(deg)	&	88.5821686	&	88.5870493		\\
$b_{{\it WISE}}$ (deg)	&	11.9790854	&	11.9269303		\\
$W1$ (mag)		&	7.952 $\pm$ 0.023	&	18.819 $\pm$ 0.536\tablenotemark{a} 	\\
$W2$ (mag)		&	7.800 $\pm$ 0.020	&	14.941 $\pm$ 0.057	\\
$W3$ (mag)		&	7.726 $\pm$ 0.017	&	$>$ 12.973		\\
$W4$ (mag)		&	7.613 $\pm$ 0.089	&	$>$ 9.07			\\
$W1-W2$ (mag)	&	0.152 $\pm$ 0.030	&	3.878 $\pm$ 0.539	\\
$W2-W3$ (mag)	&	0.074 $\pm$ 0.026	&	$<$ 1.968			\\
$ch1$ (mag)		&	...	&	15.870 $\pm$ 0.028	\\
$ch2$ (mag)		&	...	&	14.625 $\pm$ 0.020	\\
$ch1-ch2$ (mag)	&	...	&	1.245 $\pm$ 0.034	\\
$J_{MKO}$ (mag)	&	...	&	19.640 $\pm$ 0.089	\\
$H_{MKO}$ (mag)	&	...	&	19.572 $\pm$ 0.079	\\
$J_{2MASS}$ (mag)& 	8.830 $\pm$ 0.021	&	...	\\
$H_{2MASS}$ (mag)&	8.346 $\pm$ 0.021	&	...	\\
$Ks_{2MASS}$ (mag)&	8.113 $\pm$ 0.018	&	...	\\
($J-H$)$_{MKO}$ (mag)&	...	&	0.068 $\pm$ 0.119	\\
$J_{MKO}$$-W2$ (mag)&	...	&	4.699 $\pm$ 0.106	\\
$H_{MKO}$$-W2$ (mag)&	...	&	4.631 $\pm$ 0.097	\\
$\mu_{\alpha}$ (arcsec yr$^{-1}$) &	$-$1.163  $\pm$ 0.005\tablenotemark{b} 	&	$-$1.138 $\pm$ 0.102\\
$\mu_{\delta}$ (arcsec yr$^{-1}$) &		$-$0.900  $\pm$ 0.004\tablenotemark{b} 	&	$-$0.988 $\pm$ 0.106\\
$\pi$	(mas)  &	63.17 $\pm$ 3.82\tablenotemark{b} 	&	19 $\pm$ 96\\
d (pc)	&	15.83 $\pm$ 0.96	&	...	\\
$\gamma$ (km s$^{-1}$)	&	$-$34\tablenotemark{c}	&	...	\\
U (km s$^{-1}$)	\tablenotemark{d,e} &	$-101 \pm 7$	&	...	\\
V (km s$^{-1}$)	\tablenotemark{e} &	$-44 \pm 2$	&	...	\\
W (km s$^{-1}$)\tablenotemark{e} &	$33 \pm 3$		&	...	\\
$M_H$ (mag)	&	...	&	18.57 $\pm$ 0.21		\\
$M_{W2}$ (mag)&	...	&	13.94 $\pm$ 0.19		\\
T$_{eff}$ (K)	&	3483 $\pm$ 17\tablenotemark{f}	& 600 - 900\\
$[$M/H$]$	 (dex)	&	$-$0.45 $\pm$ 0.12\tablenotemark{f}	&	...		\\
$[$Fe/H$]$ (dex)	&	$-$0.64 $\pm$ 0.17\tablenotemark{f}	&	...		\\
\enddata
\tablenotetext{a}{SNR = 2, the limit for profile-fit magnitude uncertainties}
\tablenotetext{b}{\citet{vanleeuwen2007}}
\tablenotetext{c}{\citet{gizis1998}}
\tablenotetext{d}{U velocity is positive away from the Galactic center}
\tablenotetext{e}{Space motions are relative to the Sun and not the local standard of rest}								
\tablenotetext{f}{\citet{rojasayala2012}}																					
\end{deluxetable}		

\clearpage
																											
\begin{deluxetable}{llllll}						
\tabletypesize{\small}	
\tablewidth{5.8in}
\tablenum{2}							
\tablecaption{Astrometric Measurements of \W2005\label{astrometry}}
\tablehead{						
\colhead{Facility/} &					
\colhead{MJD} &						
\colhead{RA}&
\colhead{$\sigma$$_{RA}$}&
\colhead{Dec}&
\colhead{$\sigma$$_{Dec}$}\\
\colhead{Instrument} &					
\colhead{JD$-$2400000.5} &						
\colhead{(deg)}&
\colhead{(arcsec)}&
\colhead{(deg)}&
\colhead{(arcsec)}
}						
\startdata							
{\it WISE} All-Sky		&	55343.731516		&	301.3349418	&	0.378	&	+54.4094230	&	0.382\\
{\it WISE} Post-Cryo		&	55524.165		&	301.334676	&	0.38		&	+54.409242	&	0.38     \\
{\it Spitzer}/IRAC		&	55561.659165		&	301.3346646	&	0.183	&	+54.4093114	&	0.165\\
{\it Spitzer}/IRAC		&	55748.962883		&	301.3344610	&	0.198	&	+54.4091562	&	0.226\\
{\it Spitzer}/IRAC		&	55886.077227		&	301.3342873	&	0.151	&	+54.4090573	&	0.230\\	
Palomar/WIRC		&	56137.837095		&	301.3338706  	&	0.19		&	+54.4089008	&	0.22\\
Keck/NIRSPEC		&	56173.30190		&	301.33375	&	0.45		&	+54.40884	&	0.45\\
Keck/MOSFIRE		&	56211.291625		&	301.3336420	&	0.33		&	+54.4088130	&	0.31\\
Keck/MOSFIRE		&	56442.5828353	&	301.3333410	&	0.36		&	+54.4086280	&	0.29	\\
\enddata		
			
\end{deluxetable}

\clearpage
																											
\begin{deluxetable}{lcrrcc}						
\tabletypesize{\small}	
\tablenum{3}							
\tablecaption{Late-type T Dwarf $J-H$ Color Outliers \label{JHoutliers}}
\tablehead{						
\colhead{Object} &	
\colhead{Spectral Type} &					
\colhead{$J-H$} &						
\colhead{$H-ch2$}&
\colhead{Reference}	
}						
\startdata			
\multicolumn{5}{c}{Red: $J-H$ $>$ $-$0.2} \\  	
WISEJ000517.48+373720.5	&	T9	&	0.230	$\pm$	0.144	&	4.820	$\pm$	0.082	&	4\\
WISEJ032120.91$-$734758.8	&	T8	&	0.070	$\pm$	0.163	&	3.700	$\pm$	0.122	&	4\\
WISEJ041358.14$-$475039.3	&	T9	&	0.170	$\pm$	0.283	&	4.630	$\pm$	0.201	&	4\\
WISEJ054047.00+483232.4	&	T8.5	&	$-$0.130	$\pm$	0.054	&	3.850	$\pm$	0.054	&	4\\
WISEJ075946.98$-$490454.0	&	T8	&	$-$0.030	$\pm$	0.064	&	3.649	$\pm$	0.044	&	3, 4\\
WISEJ081117.81$-$805141.3	&	T9.5:	&	0.020	$\pm$	0.277	&	5.480	$\pm$	0.211	&	4\\
ULASJ095047.28+011734.3	&	T8p	&	0.020	$\pm$	0.106	&	3.650	$\pm$	0.082	&	5, 4, 6\\
WISEJ104245.23$-$384238.3	&	T8.5	&	$-$0.100	$\pm$	0.142	&	4.508	$\pm$	0.112	&	3, 4\\
WISEJ113949.24$-$332425.1	&	T7	&	0.030	$\pm$	0.106	&	2.940	$\pm$	0.082	&	2\\
WISEJ144806.48$-$253420.3	&	T8	&	0.300	$\pm$	0.170	&	3.820	$\pm$	0.122	&	2\\
WISEJ161441.46+173935.5	&	T9	&	0.613	$\pm$	0.224	&	4.253	$\pm$	0.217	&	3\\
WISEJ200520.38+542433.9	&	T8p	&	0.068	$\pm$	0.119	&	4.947	$\pm$	0.081	&	1\\
WISEJ213456.73$-$713744.5	&	T9p	&	0.100	$\pm$	0.180	&	5.742	$\pm$	0.151	&	3\\
WISEJ232519.53$-$410535.0	&	T9p	&	0.529	$\pm$	0.124	&	5.129	$\pm$	0.116	&	3\\
\multicolumn{5}{c}{Blue: $J-H$ $<$ $-$0.5} \\  	
ULASJ013939.77+004813.8   	&	T7.5	&	$-$0.690	$\pm$	0.071	&	2.790	$\pm$	0.058	&	8, 9\\
CFBDSJ030135.11$-$161418.0	&	T7p	&	$-$0.650	$\pm$	0.122	&	3.570	$\pm$	0.100	&	6, 7\\															
CFBDSJ092250.12+152741.4	&	T7	&	$-$0.530	$\pm$	0.108	&	2.700	$\pm$	0.100	&	6, 7\\
ULASJ101721.40+011817.9   	&	T8p  	&	$-$0.540	$\pm$	0.028	&	3.050	$\pm$	0.036	&	11, 9\\
WISEJ154214.00+223005.2	&	T9.5	&	$-$1.550	$\pm$	0.810	&	6.740	$\pm$	0.800	&	4\\
WISEJ161705.74+180714.1	&	T8	&	$-$0.575	$\pm$	0.112	&	4.137	$\pm$	0.080	&	10, 3\\
WISEJ180435.37+311706.4	&	T9.5:	&	$-$0.510	$\pm$	0.121	&	4.608	$\pm$	0.112	&	3\\
WISEJ181210.85+272144.3	&	T8.5:	&	$-$0.640	$\pm$	0.171	&	4.660	$\pm$	0.161	&	10, 3, 4\\
Wolf 940B	&	T8.5	&	$-$0.590	$\pm$	0.042	&	4.340	$\pm$	0.042	&	12, 13\\
\enddata	
\tablecomments{References: 1) This Paper; 2)\citet{thompson2013}; 3)\citet{kirkpatrick2011}; 4)\citet{mace2013}; 5)\citet{leggett2012}; 
6)\citet{burningham2013}; 7)\citet{albert2011}; 8)\citet{chiu2008}; 9)\citet{leggett2010a}; 10)\citet{burgasser2011}; 
11)\citet{burningham2008}; 12)\citet{burningham2009}; 13)\citet{leggett2010b}}																							
\end{deluxetable}

\clearpage

\clearpage

\clearpage

\clearpage

\clearpage

\clearpage

\clearpage

\clearpage


\clearpage

\begin{thebibliography}{}
\bibitem[Albert et al.(2011)]{albert2011} Albert, L., Artigau, 
{\'E}., Delorme, P., et al.\ 2011, \aj, 141, 203
\bibitem[Allard et al.(2003)]{allard2003} Allard, F., Guillot, T., 
Ludwig, H.-G., et al.\ 2003, Brown Dwarfs, 211, 325  
\bibitem[Baraffe et 
al.(2003)]{baraffe2003} Baraffe, I., Chabrier, G., Barman, T.~S., Allard, F., \& Hauschildt, P.~H.\ 2003, \aap, 402, 701
\bibitem[Bonfils et al.(2005)]{bonfils2005} Bonfils, X., Delfosse, X., Udry, S., et al.\ 2005, \aap, 442, 635
\bibitem[Burgasser et al.(2012)]{burgasser2012} Burgasser, A.~J., 
Gelino, C.~R., Cushing, M.~C., \& Kirkpatrick, J.~D.\ 2012, \apj, 745, 26
\bibitem[Burgasser et al.(2011)]{burgasser2011} Burgasser, A.~J., 
   Cushing, M.~C., Kirkpatrick, J.~D., et al.\ 2011, \apj, 735, 116
\bibitem[Burgasser et al.(2010)]{burgasser2010} Burgasser, A.~J., 
   Looper, D., \& Rayner, J.~T.\ 2010, \aj, 139, 2448
\bibitem[Burgasser et al.(2006a)]{burgasser2006a} Burgasser, A.~J., 
Geballe, T.~R., Leggett, S.~K., Kirkpatrick, J.~D., 
\& Golimowski, D.~A.\ 2006a, \apj, 637, 1067 
\bibitem[Burgasser et al.(2006b)]{burgasser2006BBK} Burgasser, A.~J., 
   Burrows, A., \& Kirkpatrick, J.~D.\ 2006b, \apj, 639, 1095
\bibitem[Burgasser et al.(2005)]{burgasser2005} Burgasser, A.~J., 
Kirkpatrick, J.~D., \& Lowrance, P.~J.\ 2005, \aj, 129, 2849 
\bibitem[Burgasser et al.(2002)]{burgasser2002} Burgasser, A.~J., 
Kirkpatrick, J.~D., Brown, M.~E., et al.\ 2002, \apj, 564, 421 
\bibitem[Burgasser et al.(2000)]{burgasser2000} Burgasser, A.~J., 
Kirkpatrick, J.~D., Cutri, R.~M., et al.\ 2000, ApJL, 531, L57
\bibitem[Burningham et al.(2013)]{burningham2013} Burningham, B., 
Cardoso, C.~V., Smith, L., et al.\ 2013, \mnras, 433, 457
\bibitem[Burningham et al.(2011)]{burningham2011} Burningham, B., 
Leggett, S.~K., Homeier, D., et al.\ 2011, \mnras, 414, 3590 
\bibitem[Burningham et al.(2010a)]{burningham2010a} Burningham, B., 
Leggett, S.~K., Lucas, P.~W., et al.\ 2010a, \mnras, 404, 1952 
\bibitem[Burningham et al.(2010b)]{burningham2010b} Burningham, B., 
Pinfield, D.~J., Lucas, P.~W., et al.\ 2010b, \mnras, 406, 1885
\bibitem[Burningham et al.(2009)]{burningham2009} Burningham, B., 
Pinfield, D.~J., Leggett, S.~K., et al.\ 2009, \mnras, 395, 1237 
\bibitem[Burningham et al.(2008)]{burningham2008} Burningham, B., 
Pinfield, D.~J., Leggett, S.~K., et al.\ 2008, \mnras, 391, 320
\bibitem[Burrows et al.(2011)]{burrows2011} Burrows, A., Heng, K., 
\& Nampaisarn, T.\ 2011, \apj, 736, 47 
\bibitem[Burrows et al.(2006)]{burrows2006} Burrows, A., Sudarsky, 
D., \& Hubeny, I.\ 2006, \apj, 640, 1063
\bibitem[Burrows et al.(2003)]{burrows2003} Burrows, A., Sudarsky, 
   D., \& Lunine, J.~I.\ 2003, \apj, 596, 587
\bibitem[Chiu et al.(2008)]{chiu2008} Chiu, K., Liu, M.~C., 
Jiang, L., et al.\ 2008, \mnras, 385, L53
\bibitem[Cushing et al.(2011)]{cushing2011} Cushing, M.~C., 
Kirkpatrick, J.~D., Gelino, C.~R., et al.\ 2011, \apj, 743, 50
\bibitem[Cutri et al.(2012)]{cutri2012} Cutri, R.~M., \& et al.\ 2012, VizieR Online Data Catalog, 2311, 0
\bibitem[Dawson \& De Robertis(1998)]{dawson1998} Dawson, P.~C., \& De Robertis, M.~M.\ 1998, \aj, 116, 2565
\bibitem[Day-Jones et al.(2011)]{dayjones2011} Day-Jones, A.~C., 
Pinfield, D.~J., Ruiz, M.~T., et al.\ 2011, \mnras, 410, 705
\bibitem[Deacon et al.(2012a)]{deacon2012a} Deacon, N.~R., Liu, 
M.~C., Magnier, E.~A., et al.\ 2012a, \apj, 755, 94
\bibitem[Deacon et al.(2012b)]{deacon2012b} Deacon, N.~R., Liu, 
    M.~C., Magnier, E.~A., et al.\ 2012b, \apj, 757, 100
    \bibitem[Delorme et al.(2008)]{delorme2008} Delorme, P., Willott, C.~J., Forveille, T., 
  et al.\ 2008, \aap, 484, 469
\bibitem[Dupuy \& Liu(2012)]{dupuy2012} Dupuy, T.~J., \& Liu, M.~C.\ 2012, \apjs, 201, 19
\bibitem[Fazio et al.(2004)]{fazio2004} Fazio, G.~G., Ashby, 
  M.~L.~N., Barmby, P., et al.\ 2004, \apjs, 154, 39
\bibitem[Gelino et al.(2011)]{gelino2011} Gelino, C.~R., Kirkpatrick, J.~D., Cushing, M.~C., et al.\ 2011, \aj, 142, 57
\bibitem[Giclas et al.(1968)]{giclas1968} Giclas, H.~L., Burnham, 
R., \& Thomas, N.~G.\ 1968, Lowell Observatory Bulletin, 7, 67
\bibitem[Gizis(1998)]{gizis1998} Gizis, J.~E.\ 1998, \aj, 115, 2053 
\bibitem[Gizis(1997)]{gizis1997} Gizis, J.~E.\ 1997, \aj, 113, 806
\bibitem[Gliese \& Jahrei{\ss}(1991)]{gliese1991} Gliese, W., \& Jahrei{\ss}, H.\ 1991, On: The Astronomical Data Center CD-ROM: Selected Astronomical Catalogs, Vol.~I; L.E.~Brotzmann, S.E.~Gesser (eds.), NASA/Astronomical Data Center, Goddard Space Flight Center, Greenbelt, MD,
\bibitem[Gliese(1969)]{gliese1969} Gliese, W.\ 1969, 
Veroeffentlichungen des Astronomischen Rechen-Instituts Heidelberg, 22, 1
\bibitem[Goldman et al.(2010)]{goldman2010} Goldman, B., Marsat, 
S., Henning, T., Clemens, C., \& Greiner, J.\ 2010, \mnras, 405, 1140 
\bibitem[Hansen \& Phinney(1998)]{hansen1998} Hansen, B.~M.~S., \& Phinney, E.~S.\ 1998, \mnras, 294, 557 
\bibitem[Harrington \& Dahn(1980)]{harrington1980} Harrington, R.~S., \& Dahn, C.~C.\ 1980, \aj, 85, 454 
\bibitem[Hubeny \& Burrows(2007)]{hubeny2007} Hubeny, I., \& Burrows, A.\ 2007, \apj, 669, 1248
\bibitem[Ivanova et 
al.(2013)]{ivanova2013} Ivanova, N., Justham, S., Chen, X., et al.\ 2013, \aapr, 21, 59
\bibitem[Jao et al.(2009)]{jao2009} Jao, W.-C., Mason, B.~D., 
Hartkopf, W.~I., Henry, T.~J., \& Ramos, S.~N.\ 2009, \aj, 137, 3800
\bibitem[Jenkins et al.(2009)]{jenkins2009} Jenkins, J.~S., Ramsey, 
L.~W., Jones, H.~R.~A., et al.\ 2009, \apj, 704, 975 
\bibitem[Joy(1947)]{joy1947} Joy, A.~H.\ 1947, \apj, 105, 96
\bibitem[Kimble et al.(2008)]{kimble2008} Kimble, R.~A., MacKenty, 
   J.~W., O'Connell, R.~W., \& Townsend, J.~A.\ 2008, \procspie, 7010,
\bibitem[Kirkpatrick et al.(2012)]{kirkpatrick2012} Kirkpatrick, J.~D., 
Gelino, C.~R., Cushing, M.~C., et al.\ 2012, \apj, 753, 156 
\bibitem[Kirkpatrick et al.(2011)]{kirkpatrick2011} Kirkpatrick, J.~D., 
Cushing, M.~C., Gelino, C.~R., et al.\ 2011, \apjs, 197, 19
\bibitem[Kulas et al.(2012)]{kulas2012} Kulas, K.~R., McLean, 
I.~S., \& Steidel, C.~C.\ 2012, \procspie, 8453, 
\bibitem[Larkin et al.(2006)]{larkin2006} Larkin, J., Barczys, M., 
Krabbe, A., et al.\ 2006, \procspie, 6269,
\bibitem[Lawrence et al.(2007)]{lawrence2007} Lawrence, A., Warren, 
S.~J., Almaini, O., et al.\ 2007, \mnras, 379, 1599
\bibitem[Leggett et al.(2013)]{leggett2013} Leggett, S.~K., Morley, 
C.~V., Marley, M.~S., et al.\ 2013, \apj, 763, 130
\bibitem[Leggett et al.(2012)]{leggett2012} Leggett, S.~K., Saumon, 
D., Marley, M.~S., et al.\ 2012, \apj, 748, 74
\bibitem[Leggett et al.(2010a)]{leggett2010a} Leggett, S.~K., 
Burningham, B., Saumon, D., et al.\ 2010a, \apj, 710, 1627
\bibitem[Leggett et al.(2010b)]{leggett2010b} Leggett, S.~K., Saumon, 
D., Burningham, B., et al.\ 2010b, \apj, 720, 252
\bibitem[Leggett(1992)]{leggett1992} Leggett, S.~K.\ 1992, \apjs, 
82, 351 
\bibitem[L{\'e}pine \& Bongiorno(2007)]{LB2007} L{\'e}pine, S., \& Bongiorno, B.\ 2007, \aj, 133, 889
\bibitem[L{\'e}pine et al.(2007)]{lepine2007} L{\'e}pine, S., 
Rich, R.~M., \& Shara, M.~M.\ 2007, \apj, 669, 1235
\bibitem[Liu et al.(2007)]{liu2007} Liu, M.~C., Leggett, S.~K., 
\& Chiu, K.\ 2007, \apj, 660, 1507 
\bibitem[Looper et al.(2008)]{looper2008} Looper, D.~L., Gelino, 
C.~R., Burgasser, A.~J., \& Kirkpatrick, J.~D.\ 2008, \apj, 685, 1183 
\bibitem[Lucas et al.(2008)]{lucas2008} Lucas, P.~W., Hoare, 
M.~G., Longmore, A., et al.\ 2008, \mnras, 391, 136 
\bibitem[Luhman et al.(2011)]{luhman2011} Luhman, K.~L., 
Burgasser, A.~J., \& Bochanski, J.~J.\ 2011, ApJL, 730, L9 
\bibitem[Luhman et al.(2007)]{luhman2007} Luhman, K.~L., Patten, 
B.~M., Marengo, M., et al.\ 2007, \apj, 654, 570 
\bibitem[Luyten(1976)]{luyten1976} Luyten, W.~J.\ 1976, 
Univ.~Minnesota,1976 (XXGG) (1976), 0
\bibitem[Mace et al.(2013)]{mace2013} Mace, G.~N., Kirkpatrick, 
J.~D., Cushing, M.~C., et al.\ 2013, \apjs, 205, 6
\bibitem[Mainzer et al.(2011)]{mainzer2011} Mainzer, A., Cushing, 
M.~C., Skrutskie, M., et al.\ 2011, \apj, 726, 30 
\bibitem[Mason et al.(2006)]{mason2006} Mason, B.~D., Hartkopf, 
W.~I., Wycoff, G.~L., \& Holdenried, E.~R.\ 2006, \aj, 132, 2219
\bibitem[McCaughrean et al.(2004)]{mccaughrean2004} McCaughrean, M.~J., Close, L.~M., Scholz, R.-D., et al.\ 2004, \aap, 413, 1029 
\bibitem[McLean et al.(2012)]{mclean2012} McLean, I.~S., Steidel, 
C.~C., Epps, H.~W., et al.\ 2012, \procspie, 8446,
\bibitem[McLean et al.(2007)]{mclean2007} McLean, I.~S., Prato, 
   L., McGovern, M.~R., et al.\ 2007, \apj, 658, 1217 
\bibitem[McLean et al.(2003)]{mclean2003} McLean, I.~S., McGovern, 
   M.~R., Burgasser, A.~J., Kirkpatrick, J.~D., Prato, L., 
   \& Kim, S.~S.\ 2003, \apj, 596, 561
\bibitem[McLean et al.(2000)]{mclean2000} McLean, I.~S., Graham, 
   J.~R., Becklin, E.~E., Figer, D.~F., Larkin, J.~E., Levenson, N.~A., 
   \& Teplitz, H.~I.\ 2000, \procspie, 4008, 1048
\bibitem[McLean et al.(1998)]{mclean1998} McLean, I.~S., Becklin, 
E.~E., Bendiksen, O., et al.\ 1998, \procspie, 3354, 566
\bibitem[Mugrauer et al.(2006)]{mugrauer2006} Mugrauer, M., 
Seifahrt, A., Neuh{\"a}user, R., \& Mazeh, T.\ 2006, \mnras, 373, L31 
\bibitem[Monet et al.(2003)]{monet2003} Monet, D.~G., Levine, 
S.~E., Canzian, B., et al.\ 2003, \aj, 125, 984 
\bibitem[Morley et al.(2012)]{morley2012} Morley, C.~V., Fortney, 
J.~J., Marley, M.~S., et al.\ 2012, \apj, 756, 172 
\bibitem[Murray et al.(2011)]{murray2011} Murray, D.~N., 
Burningham, B., Jones, H.~R.~A., et al.\ 2011, \mnras, 414, 575
\bibitem[Nakajima et al.(1995)]{nakajima1995} Nakajima, T., 
Oppenheimer, B.~R., Kulkarni, S.~R., et al.\ 1995, \nat, 378, 463
\bibitem[Pinfield et al.(2012)]{pinfield2012} Pinfield, D.~J., 
Burningham, B., Lodieu, N., et al.\ 2012, \mnras, 422, 1922 
\bibitem[Pinfield et al.(2008)]{pinfield2008} Pinfield, D.~J., 
Burningham, B., Tamura, M., et al.\ 2008, \mnras, 390, 304
\bibitem[Pinfield et al.(2006)]{pinfield2006} Pinfield, D.~J., 
Jones, H.~R.~A., Lucas, P.~W., et al.\ 2006, \mnras, 368, 1281 
\bibitem[Reid et al.(1995)]{reid1995} Reid, I.~N., Hawley, 
S.~L., \& Gizis, J.~E.\ 1995, \aj, 110, 1838 
\bibitem[Riaz et al.(2006)]{riaz2006} Riaz, B., Mullan, D.~J., 
\& Gizis, J.~E.\ 2006, \apj, 650, 1133
\bibitem[Rojas-Ayala et al.(2012)]{rojasayala2012} Rojas-Ayala, B., 
Covey, K.~R., Muirhead, P.~S., \& Lloyd, J.~P.\ 2012, \apj, 748, 93
\bibitem[Rosenthal et al.(1996)]{rosenthal1996} Rosenthal, E.~D., 
Gurwell, M.~A., \& Ho, P.~T.~P.\ 1996, \nat, 384, 243 
\bibitem[Ross(1939)]{ross1939} Ross, F.~E.\ 1939, \aj, 48, 163
\bibitem[Saumon \& Marley(2008)]{saumon2008} Saumon, D., \& Marley, M.~S.\ 2008, \apj, 689, 1327 
\bibitem[Saumon et al.(2007)]{saumon2007} Saumon, D., Marley, 
M.~S., Leggett, S.~K., et al.\ 2007, \apj, 656, 1136 
\bibitem[Schlaufman 
\& Laughlin(2010)]{schlaufman2010} Schlaufman, K.~C., \& Laughlin, G.\ 2010, \aap, 519, A105
\bibitem[Scholz et 
al.(2003)]{scholz2003} Scholz, R.-D., McCaughrean, M.~J., Lodieu, N., \& Kuhlbrodt, B.\ 2003, \aap, 398, L29
\bibitem[Simons \& Tokunaga(2002)]{simons2002} Simons, D.~A., \& Tokunaga, A.\ 2002, \pasp, 114, 169
\bibitem[Skrutskie et al.(2006)]{skrutskie2006} Skrutskie, M.~F., 
Cutri, R.~M., Stiening, R., et al.\ 2006, \aj, 131, 1163
\bibitem[Stauffer \& Hartmann(1986)]{stauffer1986} Stauffer, J.~R., \& Hartmann, L.~W.\ 1986, \apjs, 61, 531
\bibitem[Stephens et al.(2009)]{stephens2009} Stephens, D.~C., 
   Leggett, S.~K., Cushing, M.~C., et al.\ 2009, \apj, 702, 154 
\bibitem[Thompson et al.(2013)]{thompson2013} Thompson, M.~A., 
Kirkpatrick, J.~D., Mace, G.~N., et al.\ 2013, \pasp, 125, 809
\bibitem[Tinney et al.(2005)]{tinney2005} Tinney, C.~G., 
   Burgasser, A.~J., Kirkpatrick, J.~D.,  \& McElwain, M.~W.\ 2005, \aj, 130, 2326
   \bibitem[Tokunaga et al.(2002)]{tokunaga2002} Tokunaga, A.~T., 
   Simons, D.~A., \& Vacca, W.~D.\ 2002, \pasp, 114, 180
   \bibitem[van Altena et al.(1995)]{vanaltena1995} van Altena, W.~F., 
Lee, J.~T., \& Hoffleit, E.~D.\ 1995, New Haven, CT: Yale University Observatory, |c1995, 4th ed.
\bibitem[van Leeuwen(2007)]{vanleeuwen2007} van Leeuwen, F.\ 2007, \aap, 474, 653
\bibitem[Werner et al.(2004)]{werner2004} Werner, M.~W., Roellig, 
T.~L., Low, F.~J., et al.\ 2004, \apjs, 154, 1
\bibitem[Wilson et al.(2003)]{wilson2003} Wilson, J.~C., 
   Eikenberry, S.~S., Henderson, C.~P., et al.\ 2003, \procspie, 4841, 451
   \bibitem[Wilson et al.(2001)]{wilson2001} Wilson, J.~C., 
Kirkpatrick, J.~D., Gizis, J.~E., et al.\ 2001, \aj, 122, 1989 
   \bibitem[Wizinowich et al.(2000)]{wizinowich2000} Wizinowich, P., 
Acton, D.~S., Shelton, C., et al.\ 2000, \pasp, 112, 315 
\bibitem[Wolf(1921)]{wolf1921} Wolf, M.\ 1921, Astronomische 
Nachrichten, 213, 31 
\bibitem[Woolf et al.(2009)]{woolf2009} Woolf, V.~M., L{\'e}pine, 
S., \& Wallerstein, G.\ 2009, \pasp, 121, 117 
\bibitem[Wright et al.(2013)]{wright2013} Wright, E.~L., 
Skrutskie, M.~F., Kirkpatrick, J.~D., et al.\ 2013, \aj, 145, 84
\bibitem[Wright et al.(2010)]{wright2010} Wright, E.~L., 
Eisenhardt, P.~R.~M., Mainzer, A.~K., et al.\ 2010, \aj, 140, 1868
\bibitem[York et al.(2000)]{york2000} York, D.~G., Adelman, J., 
Anderson, J.~E., Jr., et al.\ 2000, \aj, 120, 1579
\bibitem[Young et al.(1987)]{young1987} Young, A., Sadjadi, S., 
\& Harlan, E.\ 1987, \apj, 314, 272
\end{thebibliography}
\end{document}